\documentclass[sn-mathphys]{sn-jnl}

\begin{document}

\title[Absorption and scattering of massless scalar wave from Regular Black Holes]{Absorption and scattering of massless scalar wave from Regular Black Holes}

\author*[1]{
\fnm{Mao-Yuan}
\sur{Wan}
}
\email{maoyuan.wan.physics@gmail.com}

\author[2]{
\fnm{Chen}
\sur{Wu}
}

\affil*[1]{
\orgname{University of Shanghai for Science and Technology},
\orgaddress{
\city{Shanghai},
\postcode{200093},
\country{China}
}
}

\affil[2]{
\orgname{Shanghai Advanced Research Institute, Chinese Academy of Sciences},
\orgaddress{
\city{Shanghai},
\postcode{201210},
\country{China}
}
}
 
\abstract{In this work, 
we numerically investigate the scattering and absorption cross section of the massless scalar field from some well-known regular black holes by using the partial wave approach. 
Our computational results indicate that the larger the parameters, 
the lower the associated total absorption cross section maxima. 
When compared to the Schwarzschild black hole, 
the scattering cross section is enhanced in some regular black hole spacetimes, 
meanwhile the scattering width is narrow in the forward orientation. 
Moreover, 
it is found that the null geodesics of the critical impact parameter and the geometrical optical value in the high frequency regime have similar changing behavior.
}


\maketitle

\section{Introduction} \label{sec1}

The problem of understanding how to avoid singularities in black hole spacetimes is important in general relativity. 
Bardeen \cite{bib1} generated the first regular black hole which is non-singularity during 1968. 
Such a black hole spacetime geometry without singularity that obeys the weak energy condition, 
some exotic matter or adjustment in gravity theory must be included. 
The spacetime geometry is achieved by adding a regular and bounded energy-momentum tensor that vanishes at the infinity and matches the weak energy requirement. 
Before Ay\'on-Beato et al. \cite{bib2} in 2000, 
this regular black hole spacetime was deficient in a accordant physical explanation that describing it as the gravitational field of nonlinear magnetic monopole with a mass $M$ and a charge $q$. 
Bardeen's solution has motivated deeper works about singularity avoidance may be realized generally. 
Several other scholars paid attention to theories of gravity coupled to nonlinear electrodynamics, 
and proposed other solutions in different contexts \cite{bib3,bib4,bib5,bib6,bib7,bib8,bib9}. 
The lack of singularity is associated with topological change beyond the event horizon for regular black hole solutions. 
Borde \cite{bib10} claimed that verifying the general singularity theorem is unattainable unless strong energy conditions or global hyperbolicity are presented.

In the 1970s and 1980s, 
considerable effort  was devoted to the study of the scattering and absorption of planar waves of frequency $\omega$ by black holes \cite{bib11}. 
People can regard the scattering results as a black hole fingerprint and possibly observed. 
Recently, 
these topics have been also arousing lots of interest \cite{bib12,bib13,bib14,bib15,bib16}.

The reason that this subject attracts increasing attentions are as follows: 
1)quasinormal modes  of black holes are thought to representing the poles of the corresponding black hole scattering matrix. 
2)scattering and absorption from black holes is related to many interesting phenomena, 
such as glory, 
photon orbit and superradiant scattering \cite{bib13,bib17,bib18,bib19,bib20,bib21,bib22,bib23,bib24,bib25,bib26}. 
In Ref. \cite{bib27,bib28,bib29}, 
S\'anchez noticed that the absorption cross section of the Schwarzschild black hole for massless scalar wave oscillating near the geometrical optical limit via numerical approaches.

Chen and his coworkers \cite{bib30} investigated the scattering and absorption cross section of massless scalar wave from the black hole surrounded by  magnetic field. 
Crispino et al. \cite{bib31} extended study of the scattering theory of scalar waves in the  Reissner-Nordstr\"om black hole spacetime. 
Huang et al. \cite{bib32} studied the scattering and absorption cross section of massless scalar waves by the Bardeen black hole.

In this article, 
we mainly focus on the scalar scattering process of some well-known regular black holes, 
and how different nonlinear interactions effects on scalar absorption and scattering cross sections. 
The following is the structure of this paper: 
Sect. \ref{sec2} provide a quick overview of some regular black holes and the null geodesics in these spacetimes. 
In Sect. \ref{sec3}, 
we perform an analysis of the classical dynamics of massless scalar field. 
The concentration of Sect. \ref{sec4} is the absorption and scattering cross sections of massless scalar fields for regular black holes. 
Finally, 
we come to a conclusion in Sect. \ref{sec5}. 
Throughout this paper we use natural units $c=\hbar=G=1$.

\section{Classical analysis} \label{sec2}

\subsection{The basic equations} \label{subsec2.1}

In this subsection, 
we will first give a brief introduction to the regular black holes. 
The nonlinear electrodynamic term \cite{bib1} is incorporated in the classical action to obtain the solution
\begin{equation} \label{eq1}
S=\frac{1}{16\pi G}\int\mathrm{d}^{4}\sqrt{-g}\left[R-\mathcal{L}(F)\right]
\end{equation}
where $g$ is the determinant of the metric tensor $g^{\mu\nu}$, 
$G$ is the gravitational constant, 
$R$ is the scalar curvature, 
and $\mathcal{L}(F)$ is the nonlinear electrodynamics Lagrangian $F=(F_{\mu\nu}F^{\mu\nu})/4$, 
where $F_{\mu\nu}=\bigtriangledown_{\mu}A_{\nu}-\bigtriangledown_{\nu}A_{\mu}$ denotes the intensity of the electromagnetic field.

The generic line element for spherically symmetric regular black hole solution is as described in the following
\begin{equation} \label{eq2}
ds^2 = -f(r)c^2dt^2+f^{-1}(r)dr^2+r^2(d\theta^2+\sin^2\theta d\phi^2)
\end{equation}
where $(t, r, \theta, \phi)$ are the four dimensional spacetime spherical coordinates, 
and specific forms of the lapse function $f(r)$ distinguish between the different spacetimes which relies on the energy momentum.

The lapse function $f(r)$ of the Bardeen black hole is determined by the formula in Ref. \cite{bib1}
\begin{equation} \label{eq3}
f(r)=1-\frac{2Mr^{2}}{(r^{2}+q^{2})^{\frac{3}{2}}}
\end{equation}
where $q$ and $M$ are the magnetic charge and the mass of the magnetic monopole, 
respectively. 
Ay\'on-Beato et al. \cite{bib2} interpreted this black hole as the gravitational field of a magnetic monopole arising from nonlinear electrodynamics. 
The particular nonlinear electrodynamics Lagrangian is given by $\mathcal{L}(F)=\frac{3M}{\lvert q\rvert^{3}}[\sqrt{2q^{2}F}(1+\sqrt{2q^{2}F})]^{5/2}$. 
There has horizon only if $\lvert q\rvert\le\frac{4M}{3\sqrt{3}}$. 
However the horizon is degenerate if $q=\frac{4M}{3\sqrt{3}}$ and there are no horizon if $q>\frac{4M}{3\sqrt{3}}$.

There has another renowned regular black hole is presented by Ay\'on-Beato and Garc\'{\i}a in Ref. \cite{bib3}. 
they took the nonlinear electric field as a source of charge for the solution of Maxwell field equations to obtaining this regular black hole solution. 
Its lapse function $f(r)$ is given by
\begin{equation} \label{eq4}
f(r)=1-\frac{2Mr^{2}}{(r^{2}+q^{2})^{3/2}}+\frac{q^{2}r^{2}}{(r^{2}+q^{2})^{2}}
\end{equation}
where $M$ and $q$ are the total mass and charge respectively. 
This solution is obtained from a nonlinear electrodynamics with Lagrangian density $\mathcal{L}(F)=\frac{X^{2}}{-2q^{2}}\frac{1-8X-3X^{2}}{(1-X)^{4}}-\frac{3M}{2q^{3}}\frac{X^{(15/2-5X)}}{(1-X)^{7/2}}$, 
where $X=\sqrt{-2q^{2}F}$.

Bronnikov \cite{bib4} and Berej et al. \cite{bib5} originally constructed the regular black hole by introducing the Lagrangian for nonlinear electrodynamics to first order. 
The lapse function for this black hole is given as
\begin{equation} \label{eq5}
f(r)=1-\frac{2M}{r}\left(1-\tanh\frac{r_{0}}{r}\right)
\end{equation}
where the parameter $r_{0}$ denotes a length scale defined as $r_{0}=\frac{\pi q^{2}}{8M}$ related to the electric charge.

Dymnikova \cite{bib6} presented an exact, 
regular spherically symmetric, 
charged black hole solution by using the idea proposed by Bronnikov \cite{bib4} and Berej et al. \cite{bib5}. 
This solution is constructed from a nonlinear electrodynamic theory with a Hamiltonian-like function. 
The lapse function for Dymnikova's solution is given as
\begin{equation} \label{eq6}
f(r)=1-\frac{2M}{r}\frac{2}{\pi}\left(\arctan\frac{r_{0}}{r}-\frac{r_{0}}{r^{2}+r^{2}_{0}}r\right)
\end{equation}
the parameter $r_{0}$ in the solution (\ref{eq6}) is a length scale, 
where $M$ and $q$ is the total mass and the charge respectively.

In 2006, 
Hayward \cite{bib7} found a new regular black hole spacetime that resemble the physical interpretation of the Bardeen one and has center flatness. 
This simple regular black hole implies a specific matter energy-momentum tensor that is de Sitter at the core and vanishes at large distances $r\to\infty$. 
The lapse function $f(r)$ for the Hayward black hole takes a simple form
\begin{equation} \label{eq7}
f(r)=1-\frac{2Mr^{2}}{(r^{3}+2\alpha^{2})}
\end{equation}
where $\alpha$ is a constant. 
As like the Bardeen black hole, 
this black hole can also have zero, 
one, 
or two horizons depending on the values of $M$ and $\alpha$.

Balart and Vagenas \cite{bib8} constructed charged regular black holes in the framework of Einstein-nonlinear electrodynamics theory. 
They built the general lapse function for mass distribution functions that are inspired by continuous probability distributions. 
In this paper, 
we focus on one instances of black hole solutions that use their technique. 
The lapse function is of the form
\begin{equation} \label{eq8}
f(r)=1-\frac{2M}{r}\exp\left(-\frac{q^{2}}{2Mr}\right)
\end{equation}
in the metric function $M$ and $q$ are associated with total mass and charge respectively.

All the regular black hole solutions used in this paper are summarized in Table \ref{tab1}. 
The mass $M$ of black holes is normalized to 1. 
It is necessary to emphasize that in all the mentioned black holes the lapse function $f(r)$ can have zero, 
one, 
or two horizons depending on the value of charge parameters. 
In Table \ref{tab1}, 
we list the extreme charge parameters for which the inner horizon and outer horizon coincide.

\begin{table}[t]
\begin{center}
\begin{minipage}{\textwidth}
\caption{
Summary of some well-known regular black holes in the paper. 
For more details, 
one can refer to the Refs. \cite{bib1,bib2,bib3,bib4,bib5,bib6,bib7,bib8} listed in the subsection \ref{subsec2.1}. 
When the parameter is equal to or greater than the value presented in the Extremal column, 
the event horizon degenerates or even disappears. 
In this work, the mass $M$ of black holes is normalized to 1. 
The letters A-F  represent different black holes respectively.
} \label{tab1}
\begin{tabular}{@{}llllll@{}}
\toprule
Label & Lapse function                                                                                  & Eq.         & Extremal            & Originator                & Ref.             \\
\midrule
A     & $1-\frac{2Mr^{2}}{(r^{2}+q^{2})^{3/2}}$                                                         & (\ref{eq3}) & $q\approx0.77$      & Bardeen                   & \cite{bib1,bib2} \\
B     & $1-\frac{2Mr^{2}}{(r^{2}+q^{2})^{3/2}}+\frac{q^{2}r^{2}}{(r^{2}+q^{2})^{2}}$                    & (\ref{eq4}) & $q\approx0.63$      & Ay\'on-Beato\&Garc\'{\i}a & \cite{bib3}      \\
C     & $1-\frac{2M}{r}\left(1-\tanh\frac{r_{0}}{r}\right)$                                             & (\ref{eq5}) & $r_{0}\approx0.55$  & Bronnikova                & \cite{bib4,bib5} \\
D     & $1-\frac{2M}{r}\frac{2}{\pi}\left(\arctan\frac{r_{0}}{r}-\frac{r_{0}}{r^{2}+r^{2}_{0}}r\right)$ & (\ref{eq6}) & $r_{0}\approx0.45$  & Dymnikova                 & \cite{bib6}      \\
E     & $1-\frac{2Mr^{2}}{(r^{3}+2\alpha^{2})}$                                                         & (\ref{eq7}) & $\alpha\approx1.06$ & Hayward                   & \cite{bib7}      \\
F     & $1-\frac{2M}{r}\exp\left(-\frac{q^{2}}{2Mr}\right)$                                             & (\ref{eq8}) & $q\approx1.21$      & Balart\&Vagenas           & \cite{bib8}      \\
\botrule
\end{tabular}
\end{minipage}
\end{center}
\end{table}

\subsection{Geodesics analyze} \label{subsec2.2}

We discuss the null geodesics in regular black hole spacetimes in this subsection. 
In general, 
the geodesics in regular black hole spacetimes are described using the formula \cite{bib31}
\begin{equation} \label{eq9}
\dot{s}^{2}=-f(r)\dot{t}^{2}+f^{-1}(r)\dot{r}^{2}+r^{2}(\dot{\theta}^{2}+\sin^{2}\theta\dot{\phi}^{2})=\lambda,
\end{equation}
where the over dot denotes the derivative with respect to an affine parameter $\lambda$. For massless particles we have $\lambda = 0 $.

It is useful to introduce the function $h(r)^2$
\begin{equation} \label{eq10}
h(r)^{2}=\frac{r^{2}}{f(r)},
\end{equation}
In terms of $h(r)^2$, the radius of the outermost photon sphere $r_{\rm{ph}}$ is the largest root of the equation \cite{bib33}
\begin{equation} \label{eq11}
\frac{\mathrm{d}}{\mathrm{d}r}h(r)^{2}=0,
\end{equation}

\begin{figure}[t]
\centering
\includegraphics[width=0.3\textwidth]{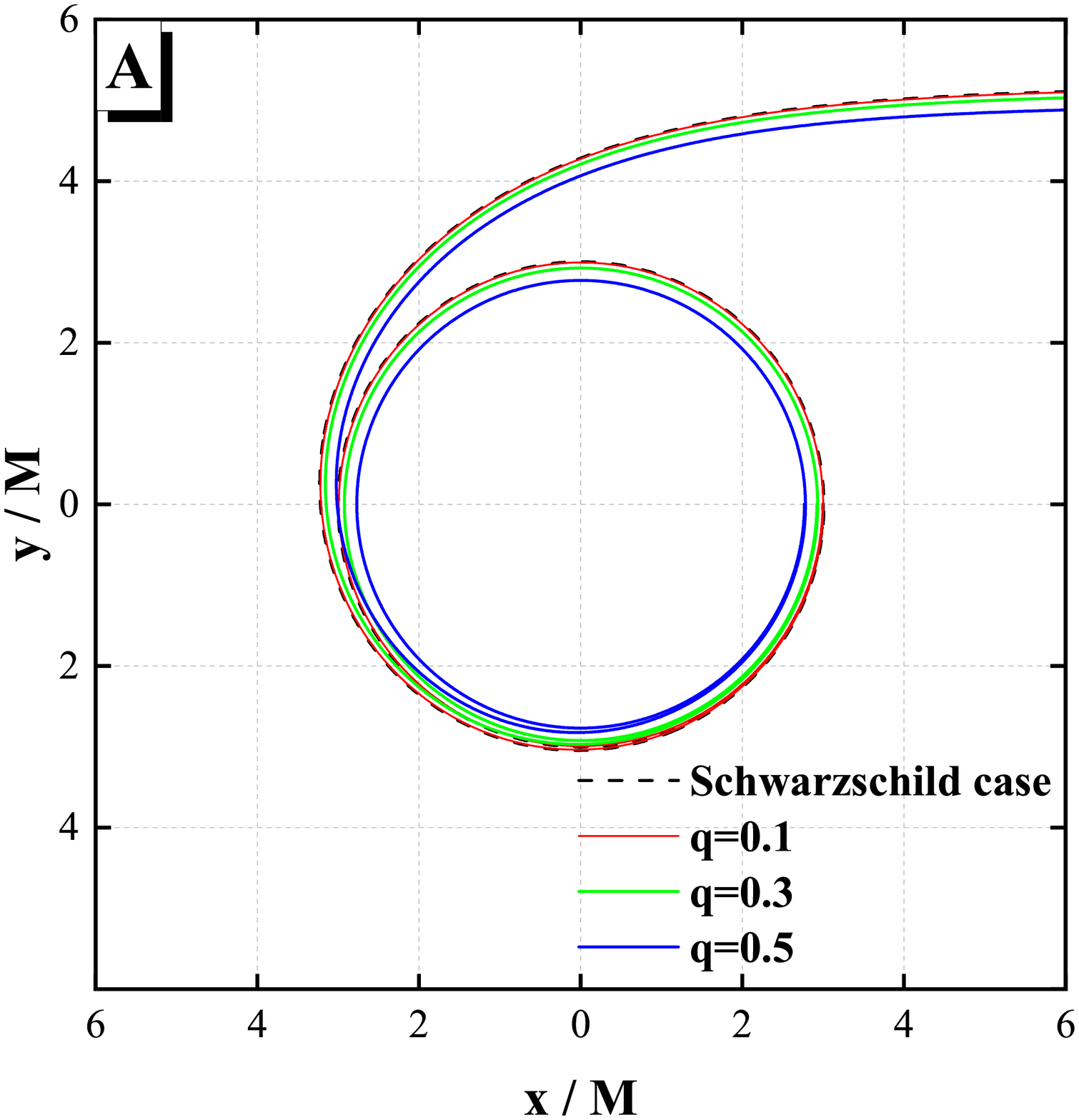}
\includegraphics[width=0.3\textwidth]{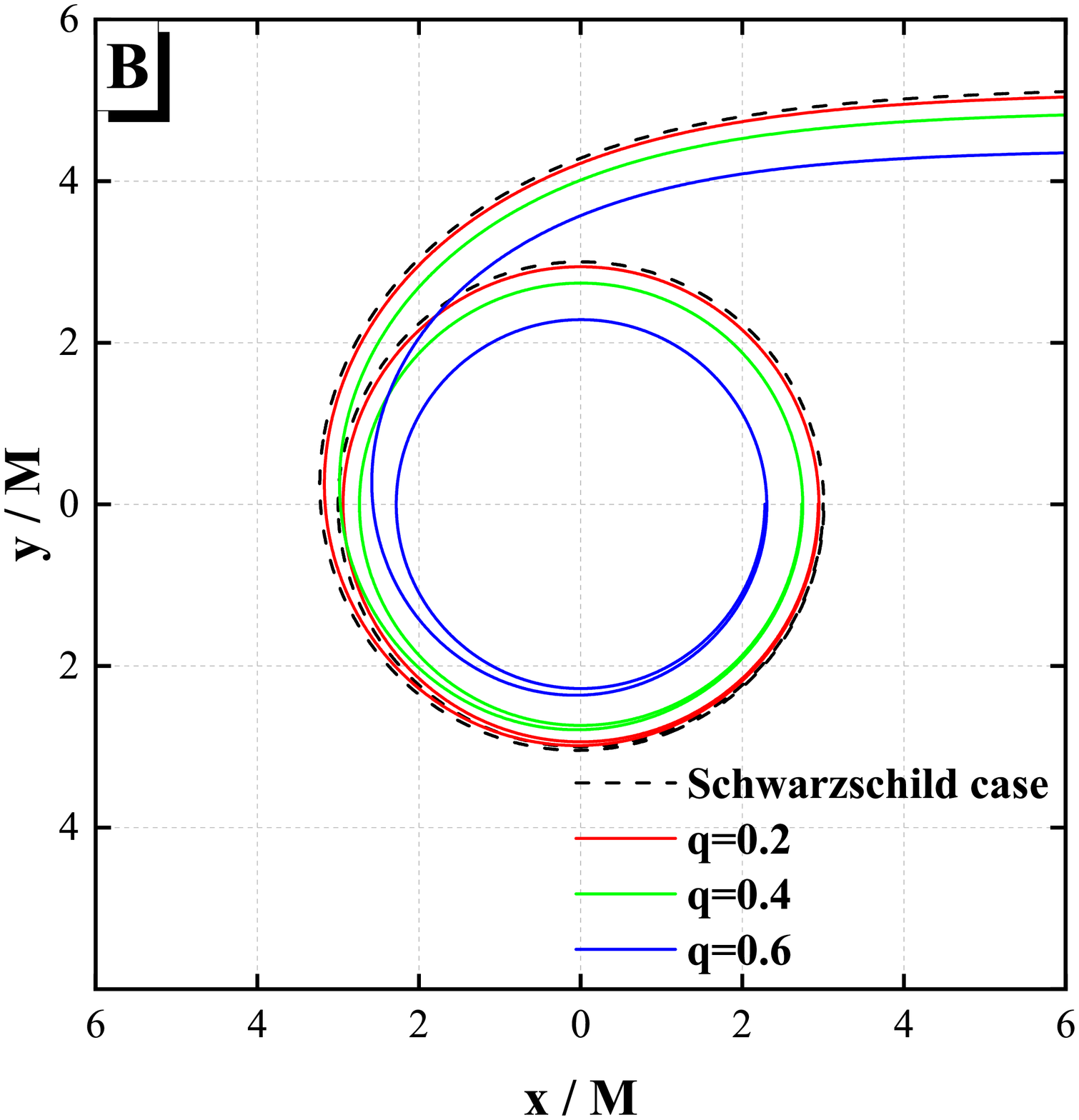}
\includegraphics[width=0.3\textwidth]{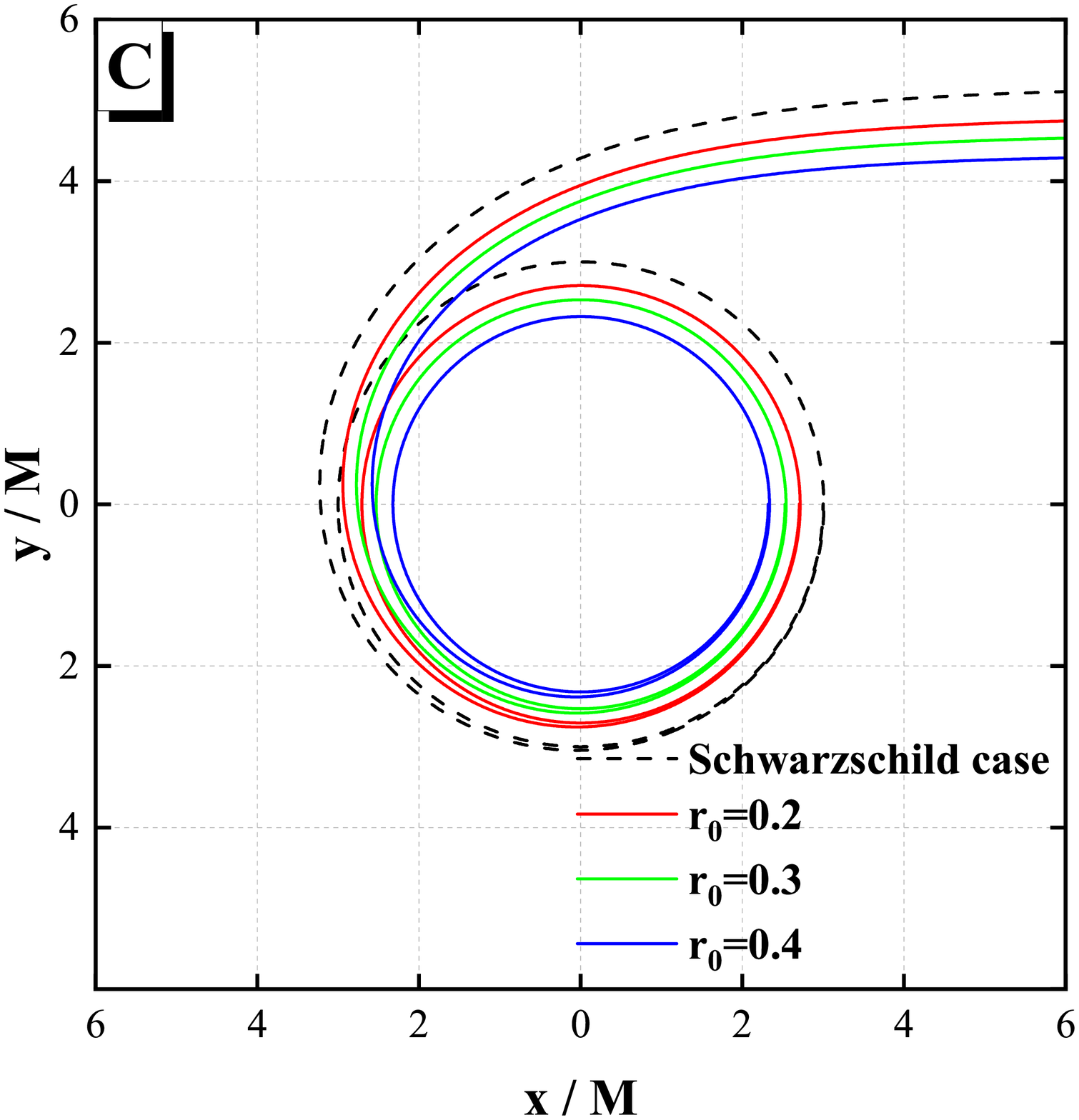}
\includegraphics[width=0.3\textwidth]{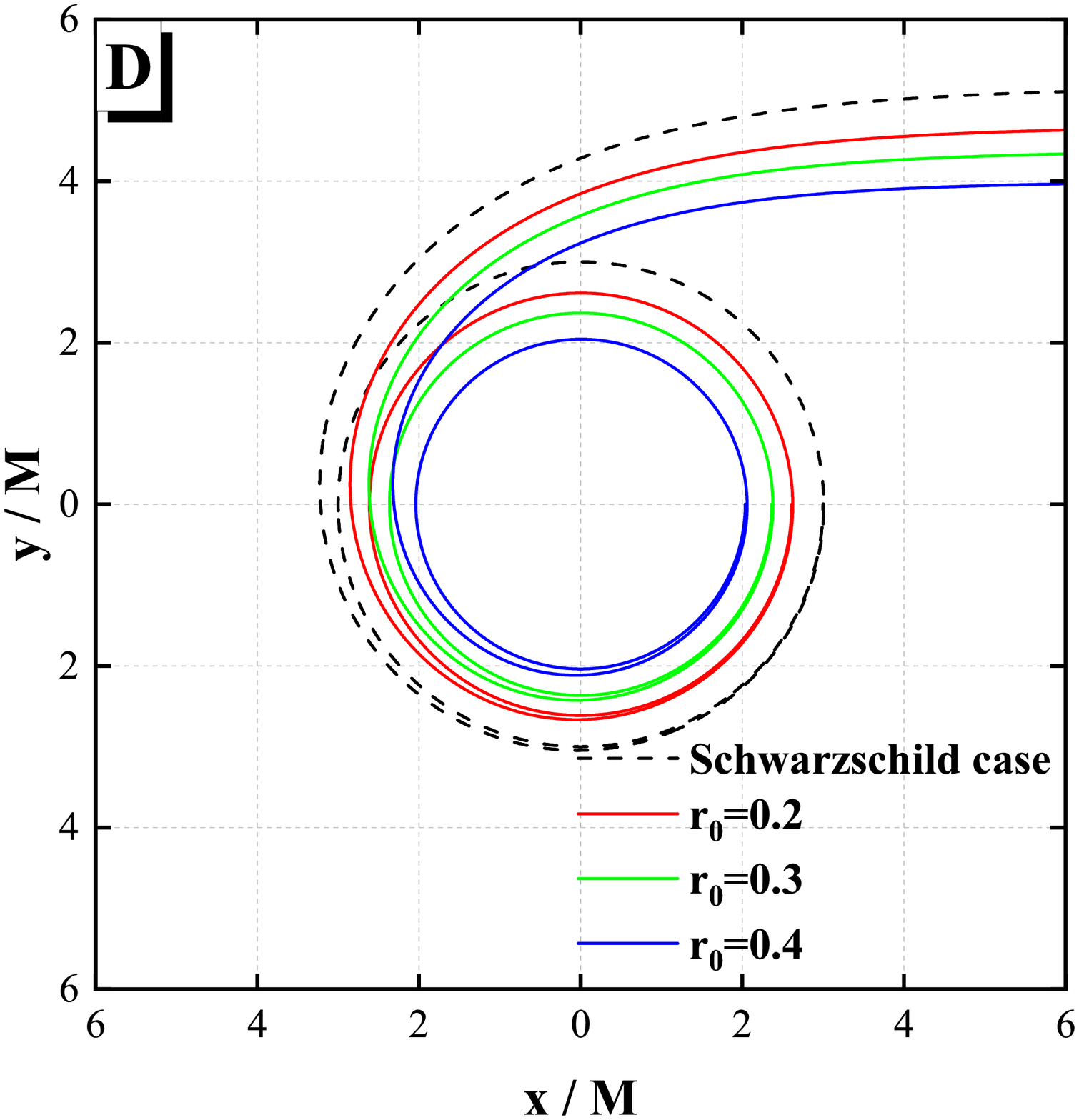}
\includegraphics[width=0.3\textwidth]{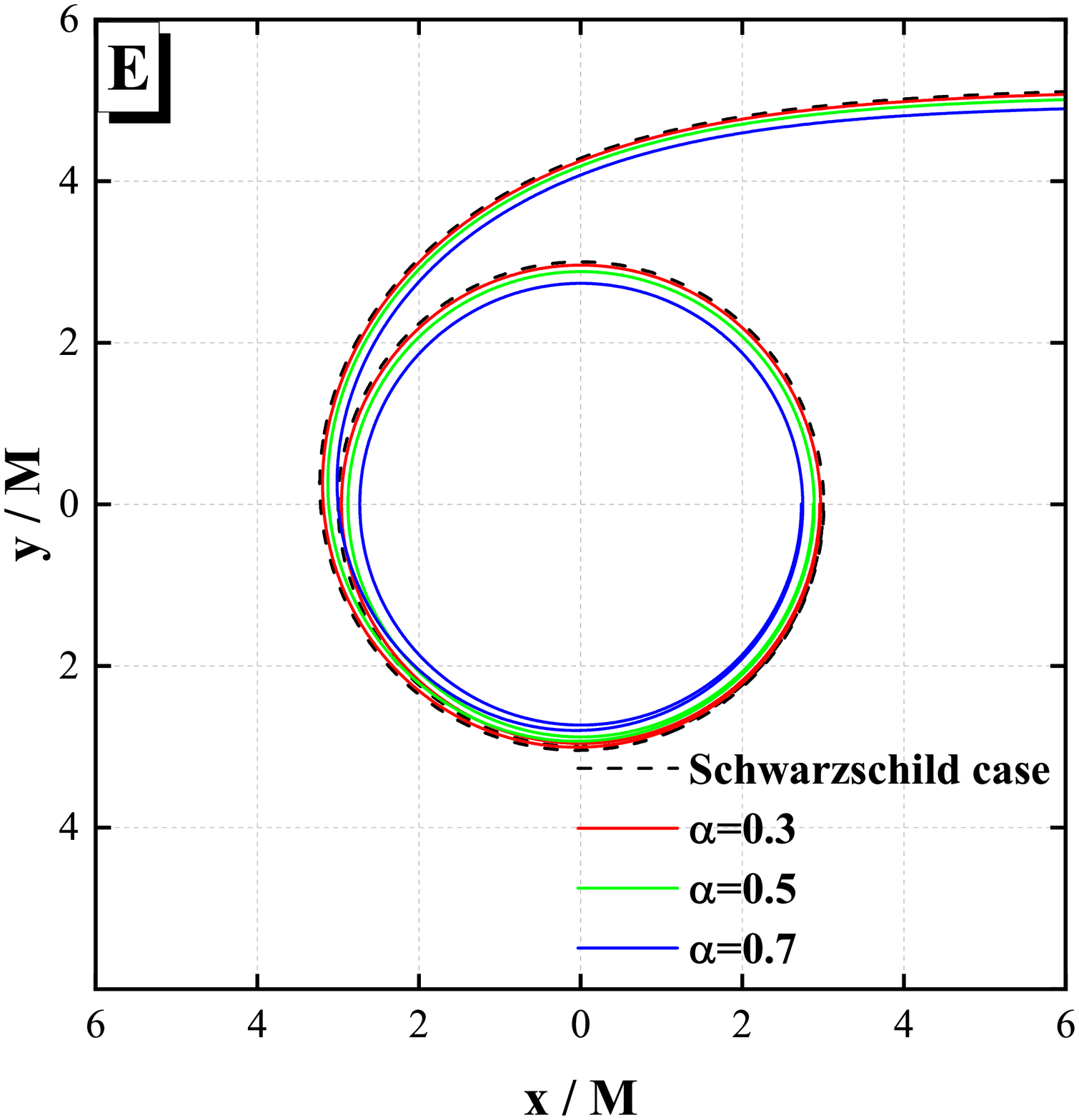}
\includegraphics[width=0.3\textwidth]{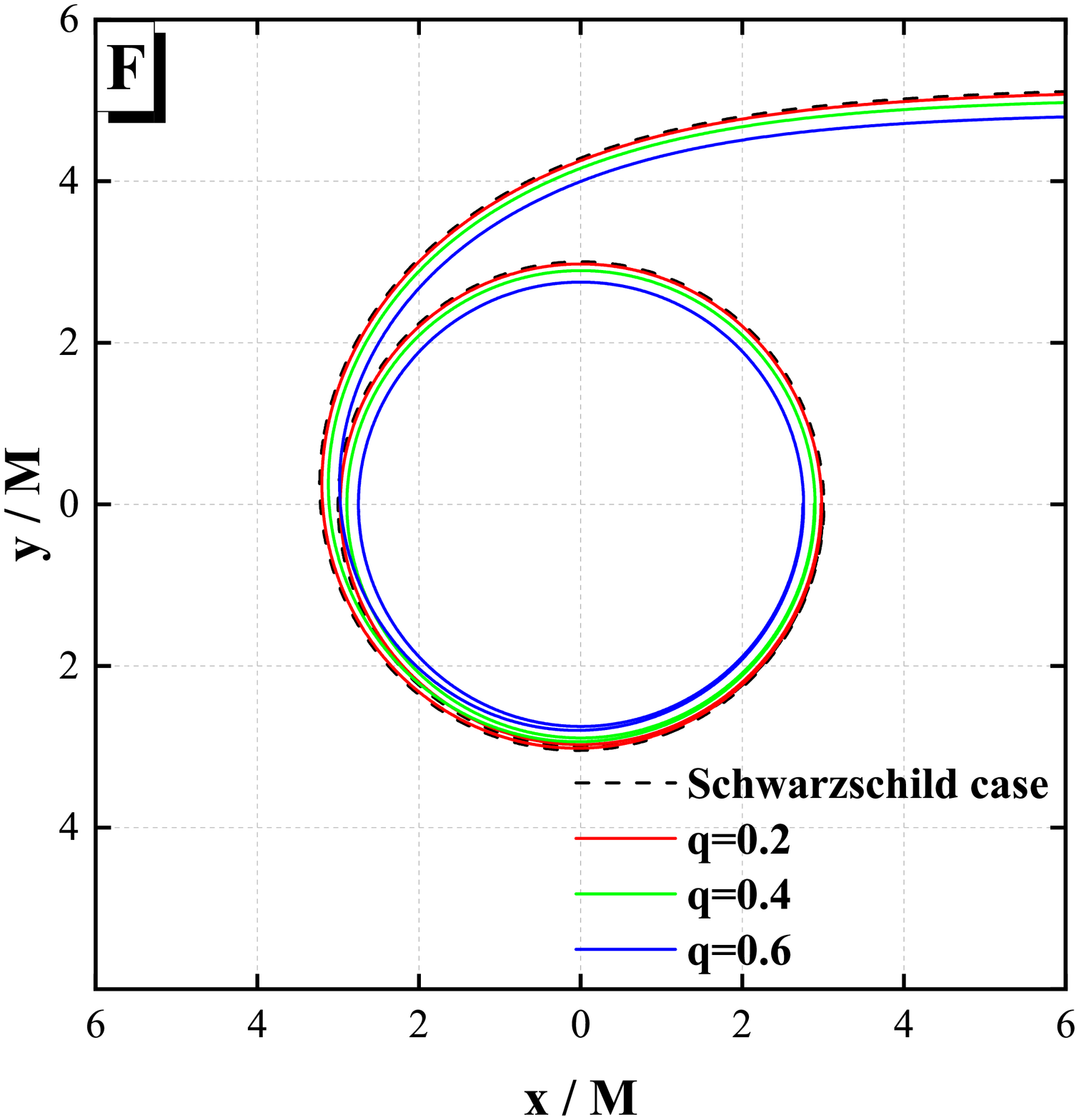}
\caption{
The null geodesics  for different critical impact parameters $b_{\rm{cr}}$ in several regular black hole spacetimes. 
We find that metric parameters for different  black hole spacetimes  considerably affect  the magnitude of the photon sphere radii. 
The letters A-F represent different black holes respectively in Table \ref{tab1}.
} \label{fig1}
\end{figure}

The critical impact parameter $b_{\rm{cr}}$ and the radius of the photon sphere $r_{\rm{ph}}$ are related by
\begin{equation} \label{eq12}
b_{\rm{cr}}=h(r_{\rm{ph}})
\end{equation}

On the equatorial plane, the orbit equation of light rays can be written as
\begin{equation} \label{eq13}
\left(\frac{\mathrm{d}r}{\mathrm{d}\varphi}\right)^{2}=\frac{r^{2}}{f^{-1}(r)}\left(\frac{h(r)^{2}}{b_{{cr}}^{2}}-1\right)
\end{equation}
In Fig. \ref{fig1}, 
we plot the null geodesics for different critical impact parameters $b_{\rm{cr}}$ in several regular black hole spacetimes. 
Then we find that metric parameters for different black hole spacetimes considerably affect the magnitude of the photon sphere radii.

Because of the existence of the innermost unstable photon sphere $r=r_{\rm{ph}}$, 
the photon can deviate from the photon sphere at any angle. 
This feature indicates that the glory phenomena will occur \cite{bib31}. 
Glory phenomena, 
like those seen in optics, 
are bright spots or halos that occur in the scattering intensity in the epipolar direction, 
the intensity and size of which depend on the wavelength of the incident perturbation. 
The approximation developed by Matzner et al. \cite{bib34} may be used to determine the magnitude and size of bright spots.

\section{Massless scalar field equation} \label{sec3}

\subsection{Effective scattering potential} \label{subsec3.1}

The dynamics of a massless  test scalar field $\Psi$ is governed by the Klein-Gordon equation, namely
\begin{equation} \label{eq14}
\frac{1}{\sqrt{-g}}\partial_{\mu}(\sqrt{-g}g^{\mu\nu}\partial_{\nu}\Psi)=0.
\end{equation}
Regular black holes are spherically symmetric spacetime with a global timelike Killing vector field, 
$\partial_{t}$, 
for $r>r_{\rm{EH}}$ \cite{bib31}, 
where $r_{\rm{EH}}$ is the outer event horizon. 
As a result, we can separate the variables
\begin{equation} \label{eq15}
\Psi=\frac{\psi_{\omega l}(r)}{r}Y_{lm}(\theta,\phi)e^{-i\omega t}
\end{equation}
$Y_{lm}(\theta,\phi)$ denotes a scalar spherical harmonic function. 
The radially  Schr\"odinger-like equation for the massless scalar field may therefore be obtained
\begin{equation} \label{eq16}
\left[f(r)\frac{\mathrm{d}}{\mathrm{d}r}f(r)\frac{\mathrm{d}}{\mathrm{d}r}+\omega^{2}-V_{\rm{eff}}(r)\right]\psi_{\omega l}(r)=0
\end{equation}
$V_{\rm{eff}}(r)$ denotes the effective scattering potential:
\begin{equation} \label{eq17}
V_{\rm{eff}}(r)=f(r)\left[\frac{1}{r}\frac{\mathrm{d}f(r)}{\mathrm{d}r}+\frac{l(l+1)}{r^2}\right]
\end{equation}

We present Regge-Wheeler type coordinates $r_{*}$, also known as Tortoise coordinates, 
for a better study of the asymptotic infinity limit of radial equation (\ref{eq16}). 
If we introduce the tortoise coordinate
\begin{equation} \label{eq18}
r_{*}=\int f(r)^{-1}\mathrm{d}r,
\end{equation}
the semi-infinite interval $r\in(2M,+\infty)$ is mapped to the infinite interval $r_{*}\in(-\infty,+\infty)$, 
and the effective scattering potential $V_{\rm{eff}}(r)$ changes into $V_{\rm{eff}}(r_*)$. 
After introducing this coordinate transition, 
the equation of motion of scalar field  has a Schr\"odinger-like form
\begin{equation} \label{eq19}
\left[\frac{\mathrm{d}^{2}}{\mathrm{d}r_{*}^{2}}+\omega^{2}- V_{\rm{eff}}(r_{*})\right]\psi_{\omega l}(r_{*}) =0
\end{equation}

\begin{figure}[t]
\centering
\includegraphics[width=0.3\textwidth]{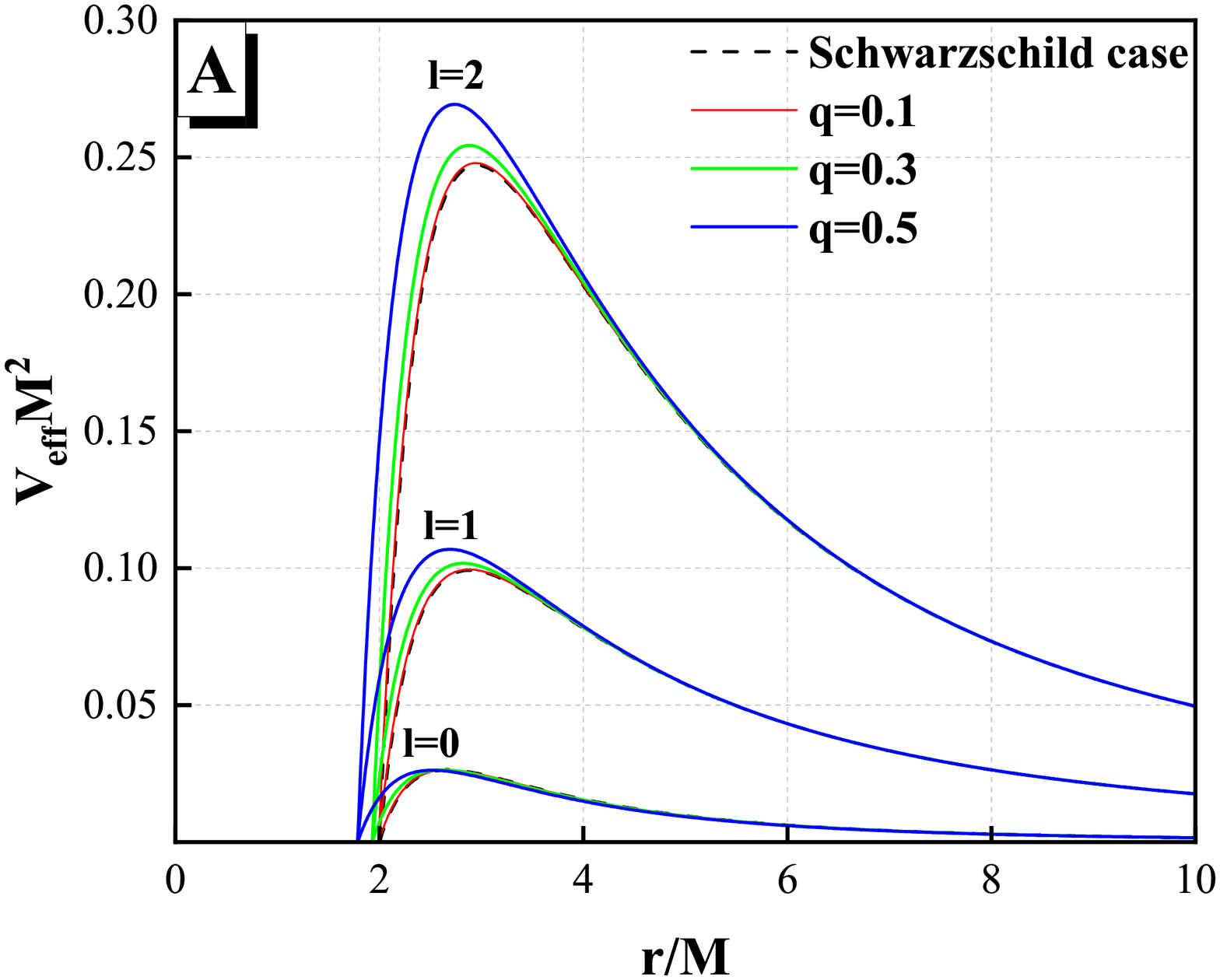}
\includegraphics[width=0.3\textwidth]{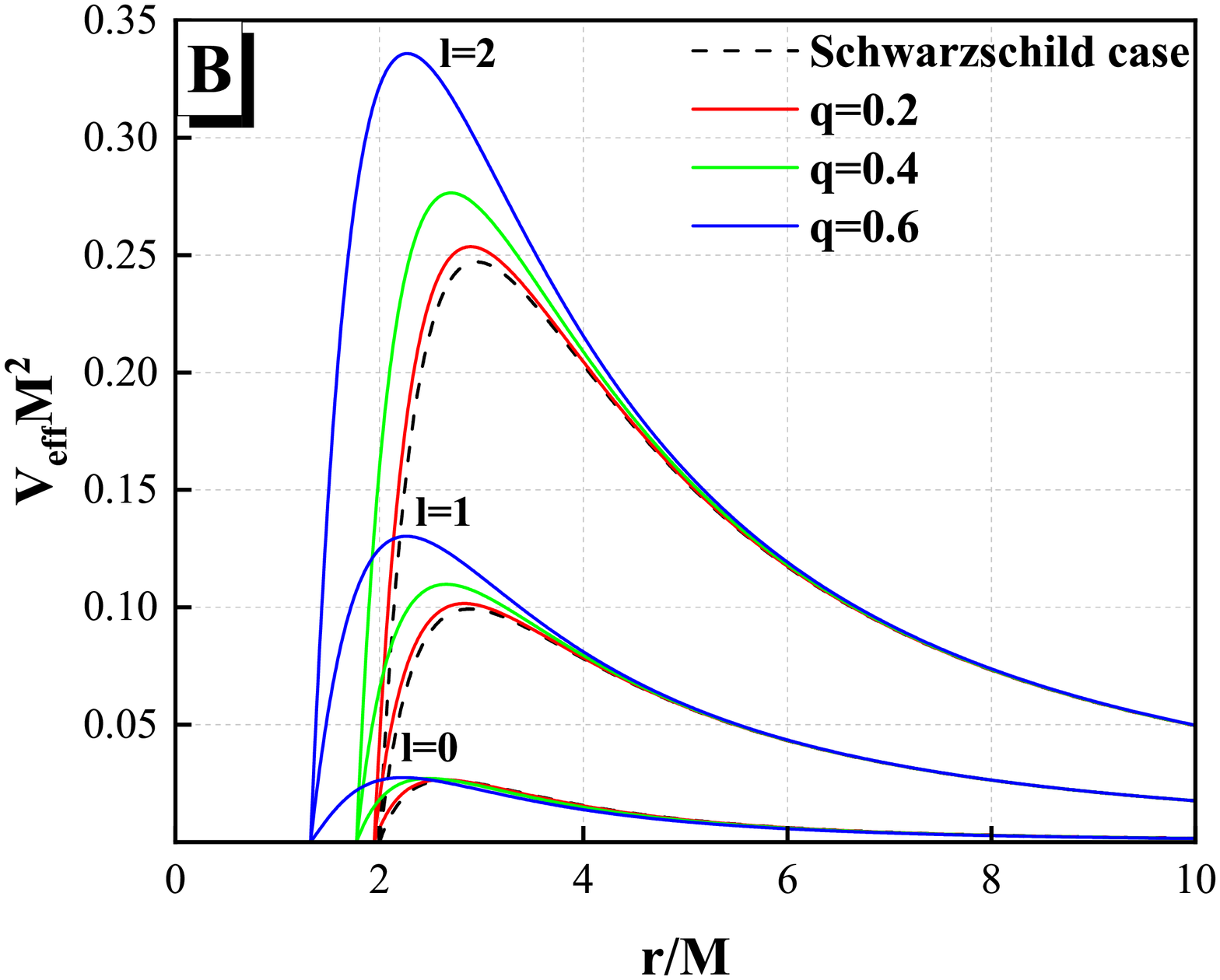}
\includegraphics[width=0.3\textwidth]{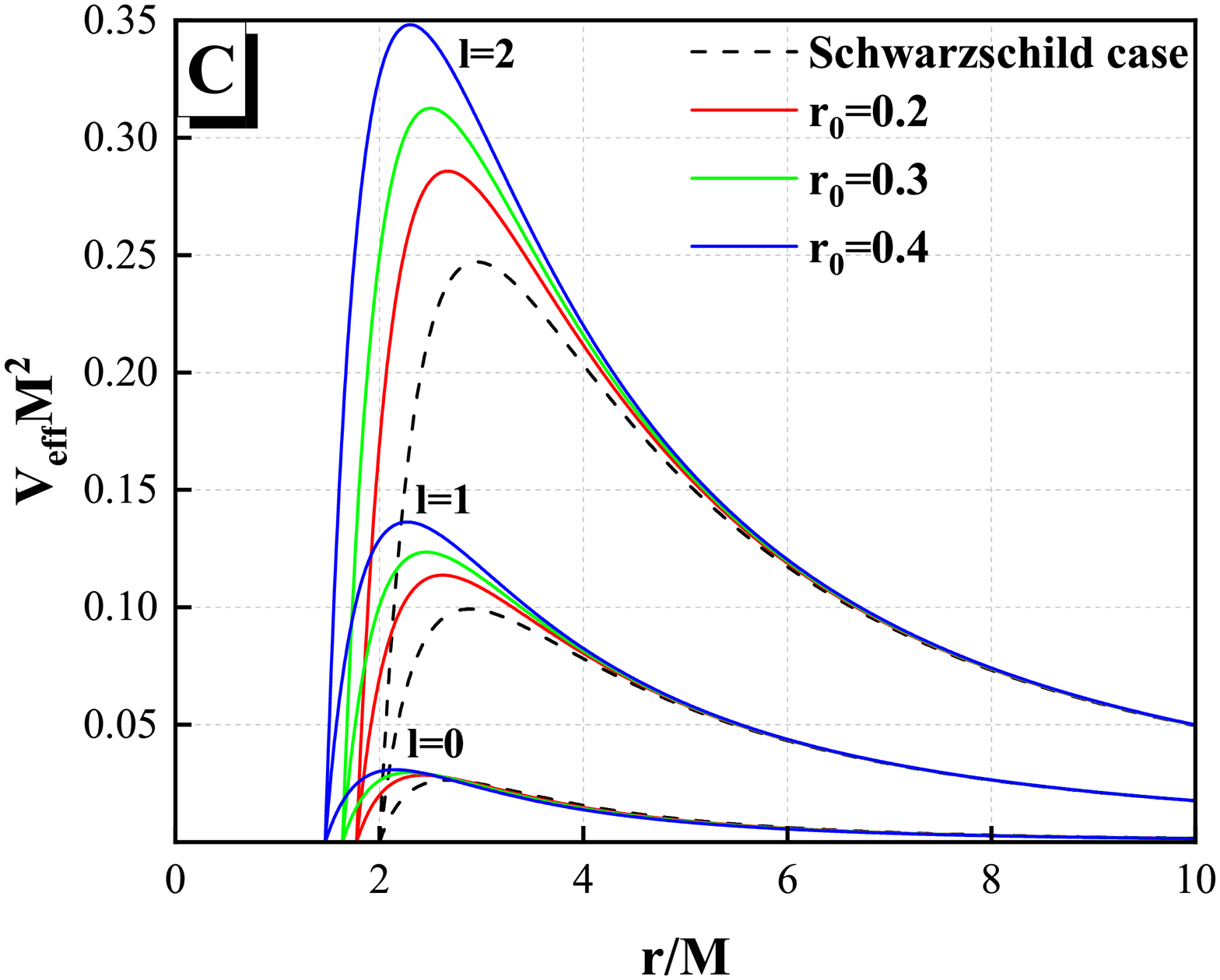}
\includegraphics[width=0.3\textwidth]{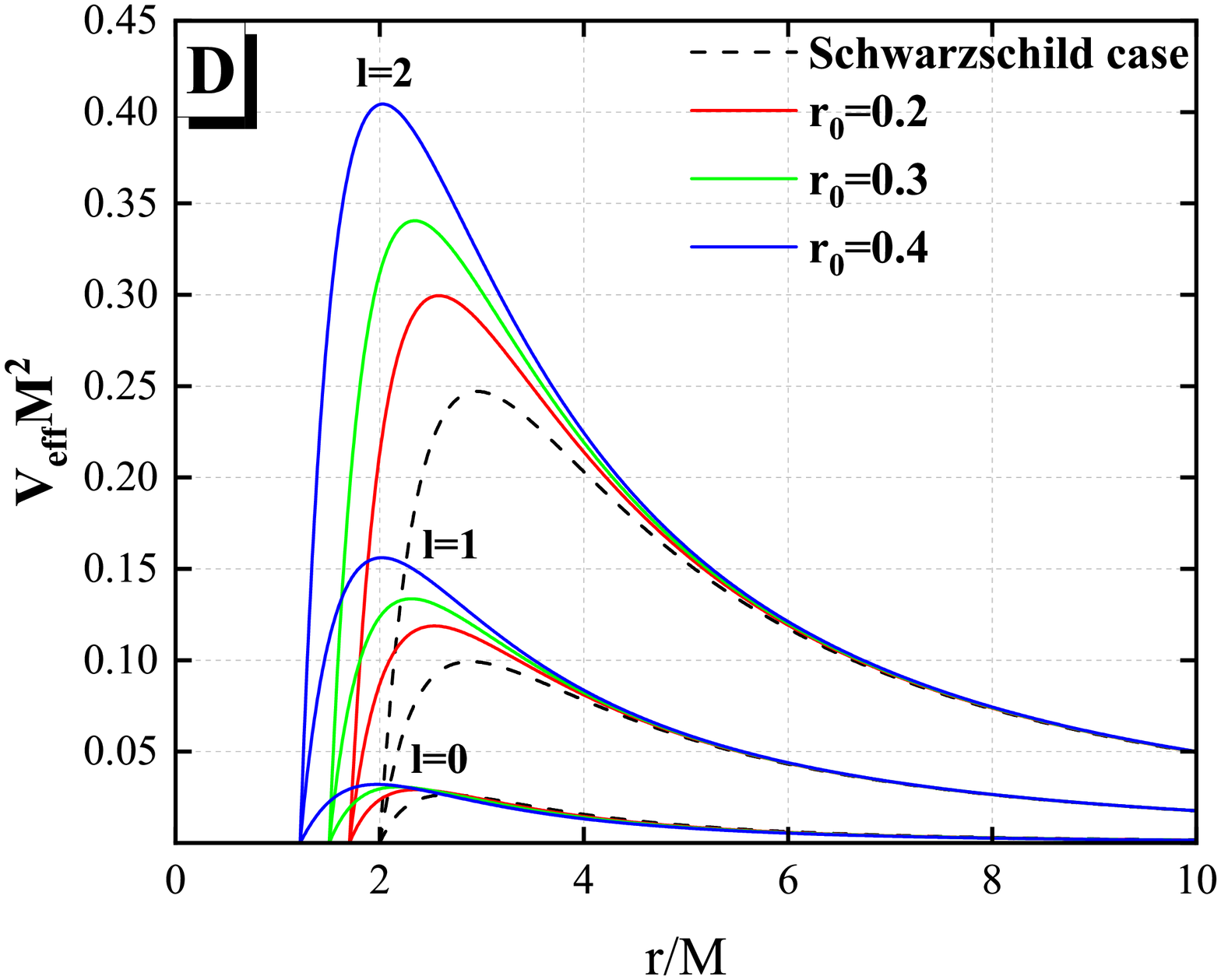}
\includegraphics[width=0.3\textwidth]{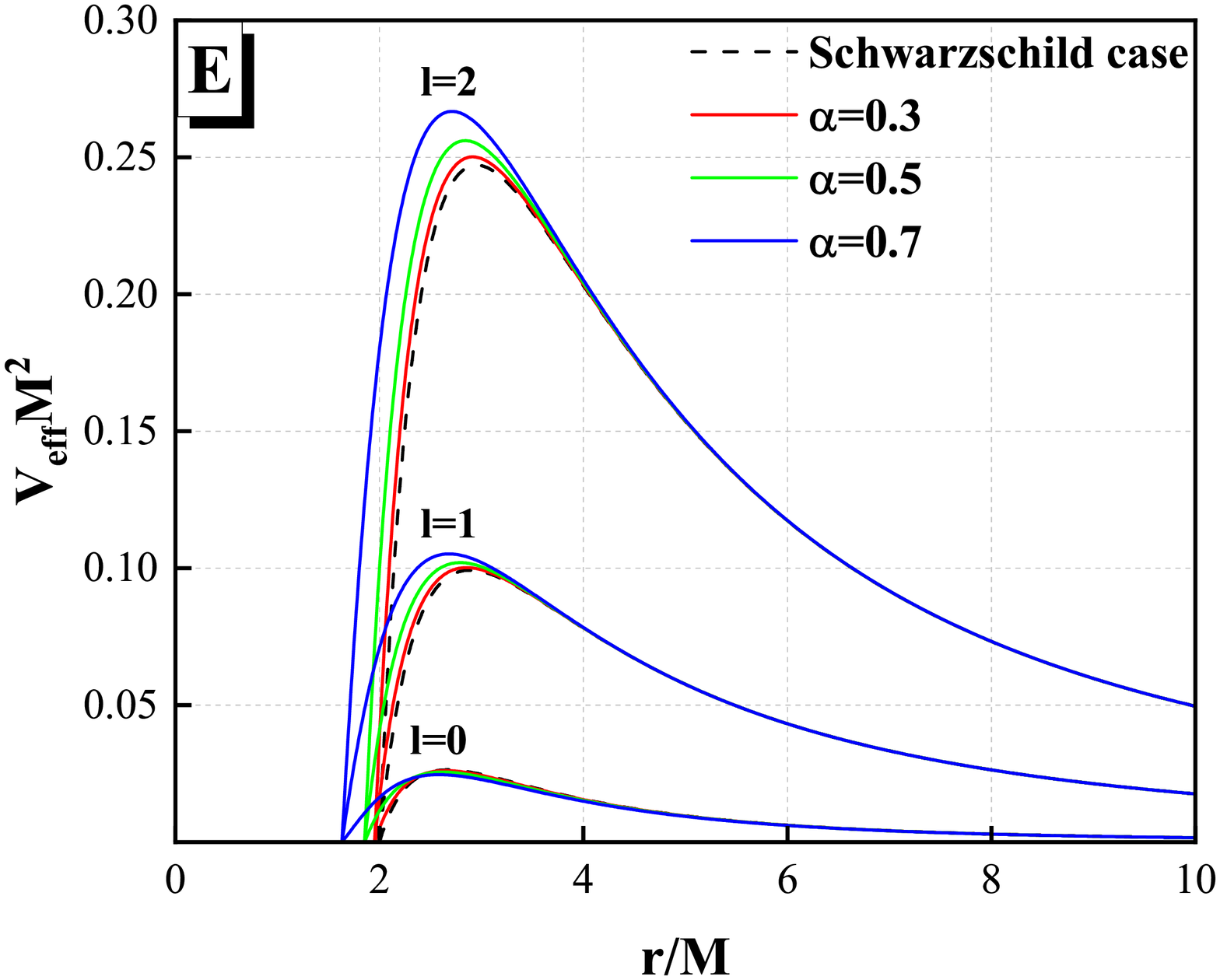}
\includegraphics[width=0.3\textwidth]{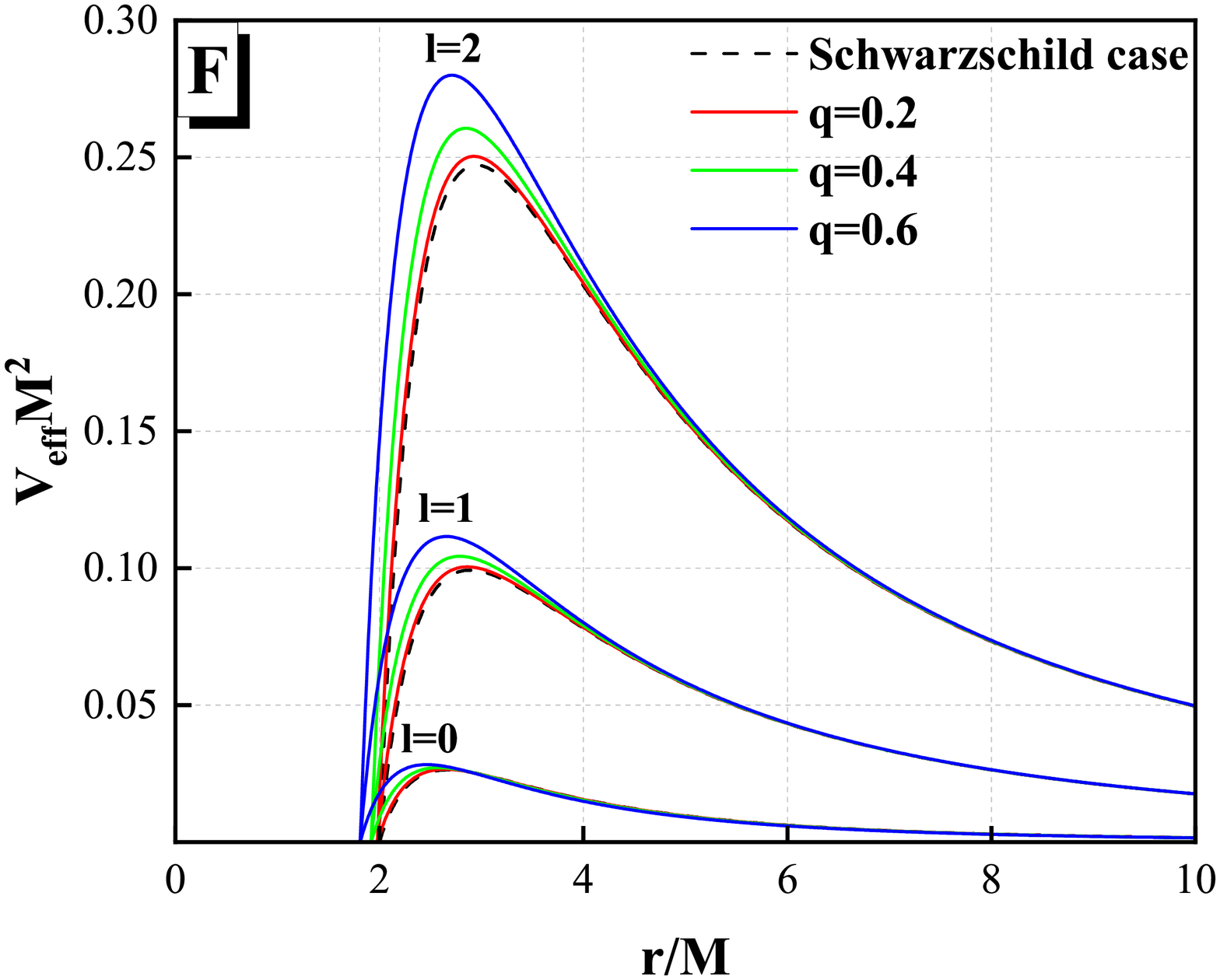}
\caption{
The effective scattering potentials $V_{\rm{eff}}$ as a function of $r$ for different choices of $l$ in the background of different regular black holes. 
The larger the value of $l$ is, the higher the scattering barrier is. 
The mass $M$ is normalized to 1. 
The letters A-F represent different black holes respectively in Table \ref{tab1}.
} \label{fig2}
\end{figure}

In Fig. \ref{fig2} ,
the effective potential $V_{\rm{eff}}(r)$ is plotted with $l=0, 1, 2$. 
From this figure, 
we can see that the peak  value of potential barrier gets upper when the angular quantum number $l$ increases. 
When the value of $l$ is fixed, 
the height of the scattering barrier increases as the metric parameter  increases. 
The results also reveal that as the radial distance increases, 
the numerical difference between the curves diminishes.

For our scattering problem, 
we consider plane scalar waves incoming from the infinite null past. 
Therefore, 
we are interested in solutions of equation (\ref{eq19}) subjected to the following boundary conditions
\begin{equation} \label{eq20}
\psi_{\omega l}(r_{*})\sim
\begin{cases}
A_{\rm{in}}e^{-i \omega r_{*}}+A_{\rm{out}}e^{i \omega r_{*}} & \rm{for\quad} r_{*}\to+\infty \\
e^{-i \omega r_{*}}                                               & \rm{for\quad} r_{*}\to-\infty \\
\end{cases}
\end{equation}
with the relation $1+\lvert A_{\rm{out}}\rvert^{2}=\lvert A_{\rm{in}}\rvert^{2}$ satisfied. 
The phase shifts of the scattered waves are defined by
\begin{equation} \label{eq21}
e^{2i\delta_{l}}=(-1)^{l+1}\frac{A_{\rm{out}}}{A_{\rm{in}}}
\end{equation}

Based on the quantum mechanics, when an incident wave travels through the potential barrier, 
part of the wave is reflected back and part of the wave goes through the potential barrier. 
By using the the conservation of the flux, 
the transmission amplitude $T_{l}$ and reflection amplitude $R_{l}$ satisfy
\begin{equation} \label{eq22}
\lvert T_{l}\rvert^{2}+\lvert R_{l}\rvert^{2}=1,
\end{equation}
where $T_{l}$ and $R_{l}$ are defined by $T_{l}=1/A_{in}$ and $R_{l}=A_{out}/A_{in}$, 
respectively.

\subsection{The Mashhoon method} \label{subsec3.2}

To calculate the absorption cross section and scattering cross section of the scalar field scattered by regular black holes in this study, 
we need to solve the radial equation under proper boundary conditions to obtain the phase shift by using the Numerov numerical method and the P\"oschl-Teller potential approximation \cite{bib35}.

We apply the P\"oschl-Teller potential $V_{\rm{PT}}$ to approximate the effective scattering potential in order to obtain analytic results. 
The potential of P\"oschl-Teller is formulated as
\begin{equation} \label{eq23}
V_{PT}(r_{*})=\frac{V_{0}}{\cosh^{2}[\kappa(r_{*}-r_{*,\rm{max}})]}
\end{equation}
where $V_{0}$ and $-2V_{0}\kappa^{2}$ denote the peak value of the effective potential and the function curvature when the effective potential is at its highest value $r_{*}=r_{*,\rm{max}}$, 
respectively. 
More specifically
\begin{equation} \label{eq24}
\kappa^{2}=-\frac{1}{2V_{0}}\left.\frac{\mathrm{d}^{2}V_{eff}}{\mathrm{d}r_{*}^{2}}\right\rvert_{r_{*}=r_{*,\text{max}}}=-\frac{1}{2V_{0}}f(r)\left.\frac{\mathrm{d}}{\mathrm{d}r}\left[f(r)\frac{\mathrm{d}V_{eff}}{\mathrm{d}r}\right]\right\rvert_{r=r_{\text{max}}}
\end{equation}
The potential $V_{\rm{PT}}$ can be adjusted to matching its maximum of the effective potential with parameters.

One can analytically obtain the radial wave function by replacing $V_{\rm{eff}}(r_{*})$ with the P\"oschl-Teller potential $V_{PT}(r_{*})$. 
As a consequence, 
the transmission coefficient $T_{l}$ and reflection coefficient $R_{l}$ are given by \cite{bib36}
\begin{equation} \label{eq25}
\begin{aligned}
T_{l} & =\frac{\Gamma(\frac{-i\omega}{\kappa})\Gamma(1+\beta+\frac{i\omega}{\kappa})\Gamma(-\beta+\frac{i\omega}{\kappa})}{\Gamma(\frac{i\omega}{\kappa})\Gamma(1+\beta)\Gamma(-\beta)} \\
\\
R_{l} & =\frac{\Gamma(1+\beta+\frac{i\omega}{\kappa})\Gamma(-\beta+\frac{i\omega}{\kappa})}{\Gamma(1+\frac{i\omega}{\kappa})\Gamma(\frac{i\omega}{\kappa})}                             \\
\end{aligned}
\end{equation}
where
\begin{equation} \label{eq26}
\beta=-\frac{1}{2}+\left[\frac{1}{4}-\frac{V_{0}}{\kappa^{2}}\right]^{\frac{1}{2}}
\end{equation}

\section{Numerical results} \label{sec4}

\subsection{Absorption cross section} \label{subsec4.1}

It is well known that the total absorption cross section can be calculated as
\begin{equation} \label{eq27}
\sigma_{\rm{abs}}=\sum\limits_{l=0}^{\infty}\sigma^{(l)}_{\rm{abs}}
\end{equation}
where $\sigma^{(l)}_{\rm{abs}}$ denotes the partial absorption cross section and can be expressed by transmission coefficient:
\begin{equation} \label{eq28}
\sigma^{(l)}_{\text{abs}}=\frac{\pi}{\omega^{2}}(2l+1)(1-\lvert e^{2i\delta_{l}}\rvert^{2})=\frac{\pi}{\omega^{2}}(2l+1)\lvert T_{\omega l}\rvert^{2}
\end{equation}

\begin{figure}[t]
\centering
\includegraphics[width=0.3\textwidth]{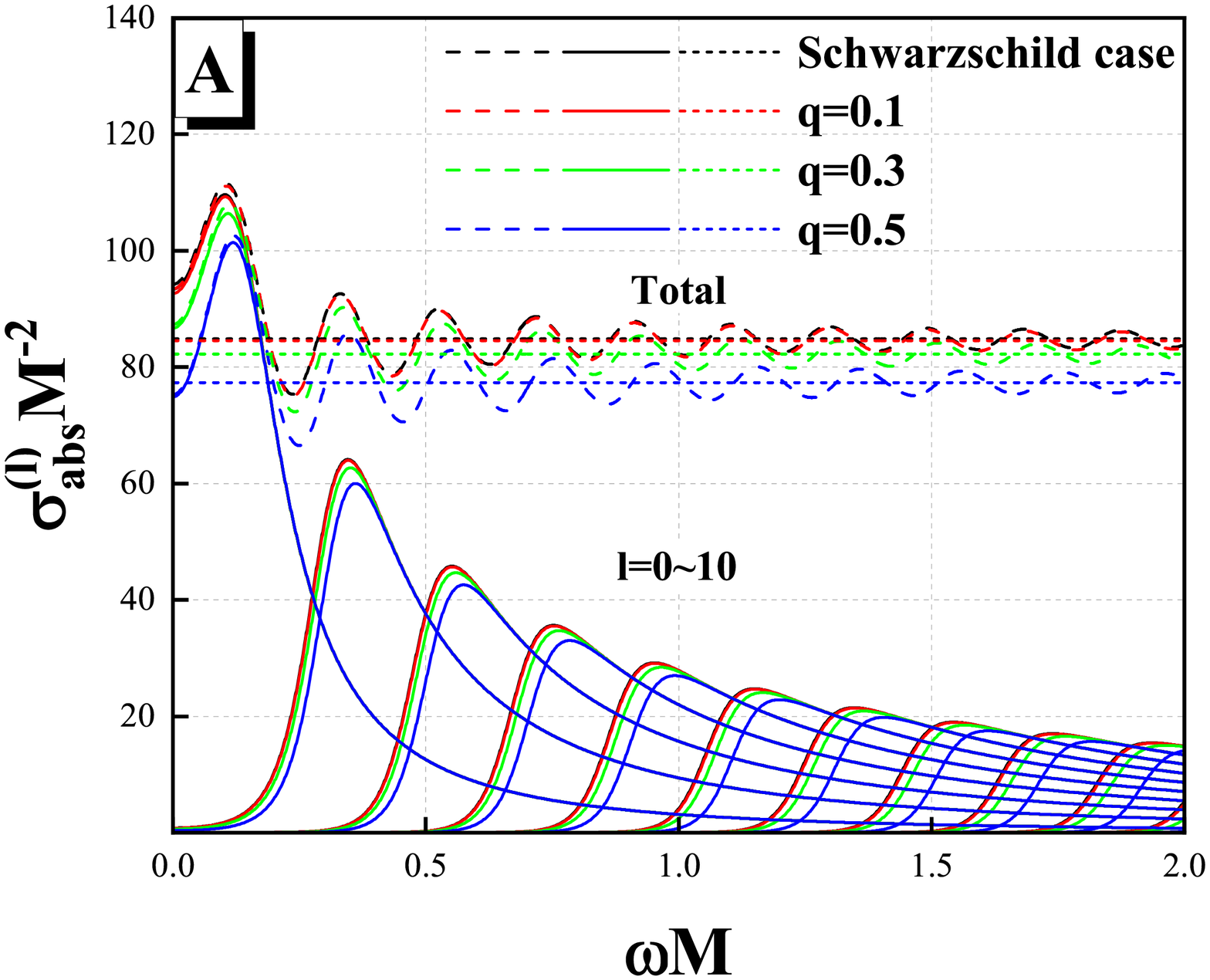}
\includegraphics[width=0.3\textwidth]{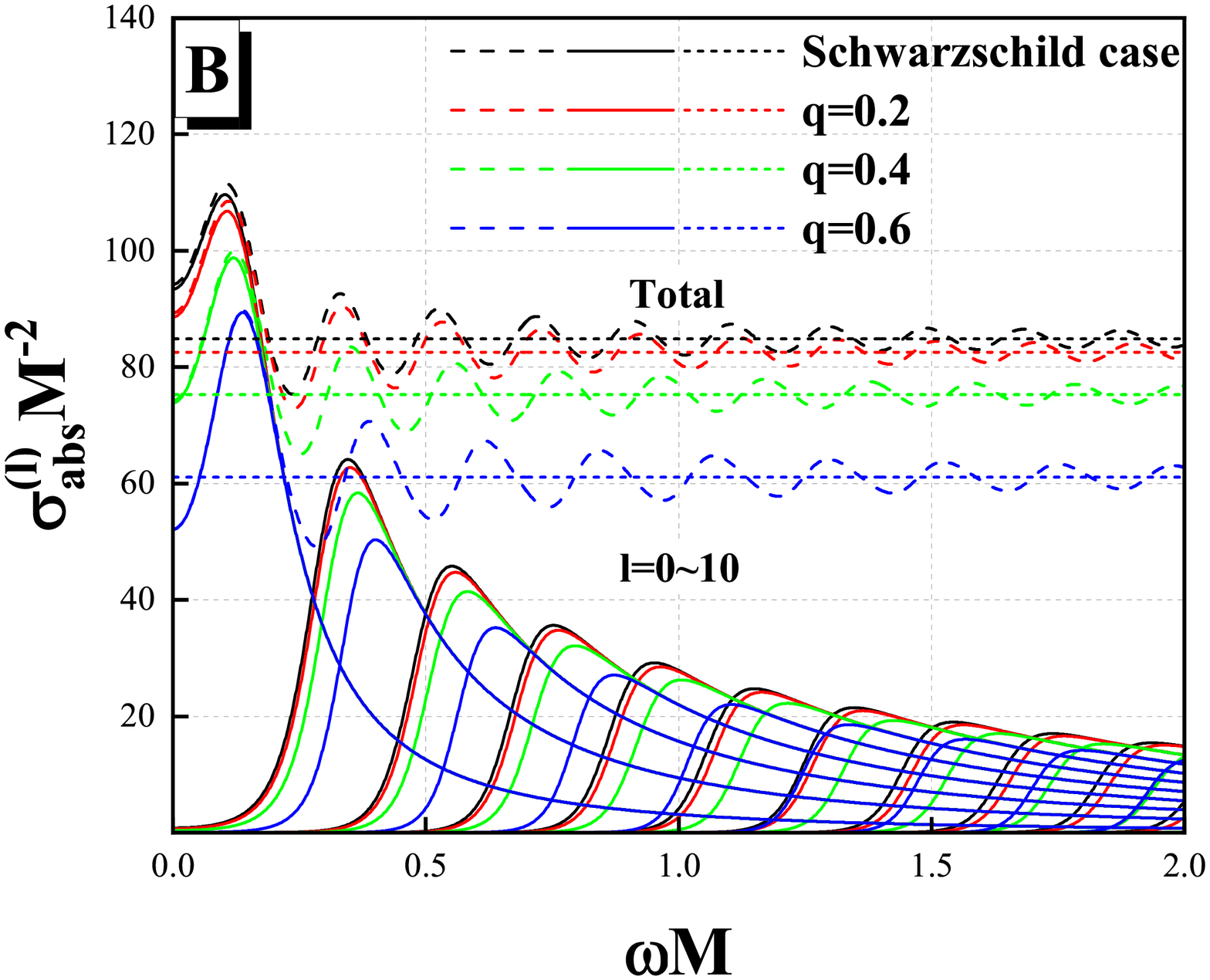}
\includegraphics[width=0.3\textwidth]{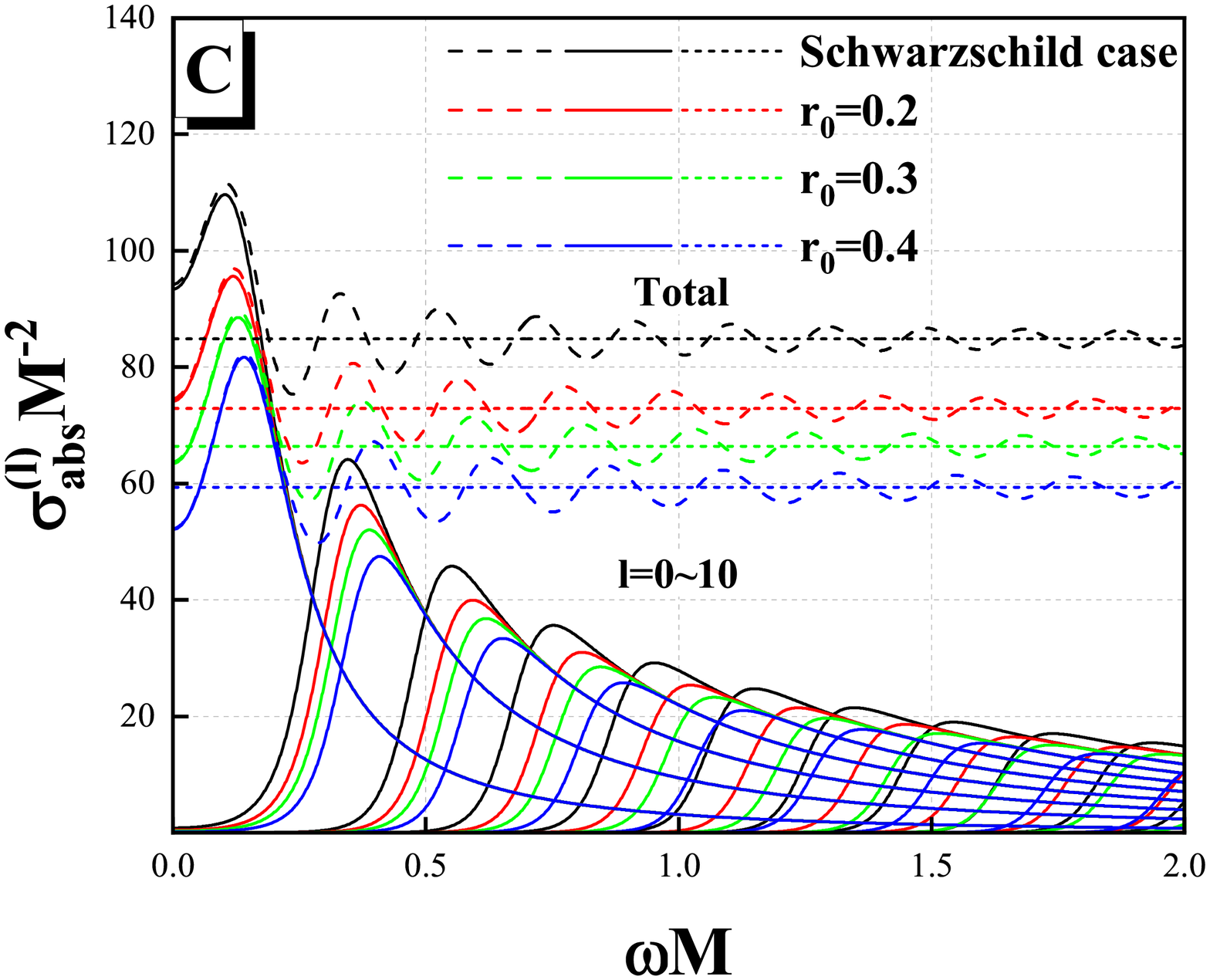}
\includegraphics[width=0.3\textwidth]{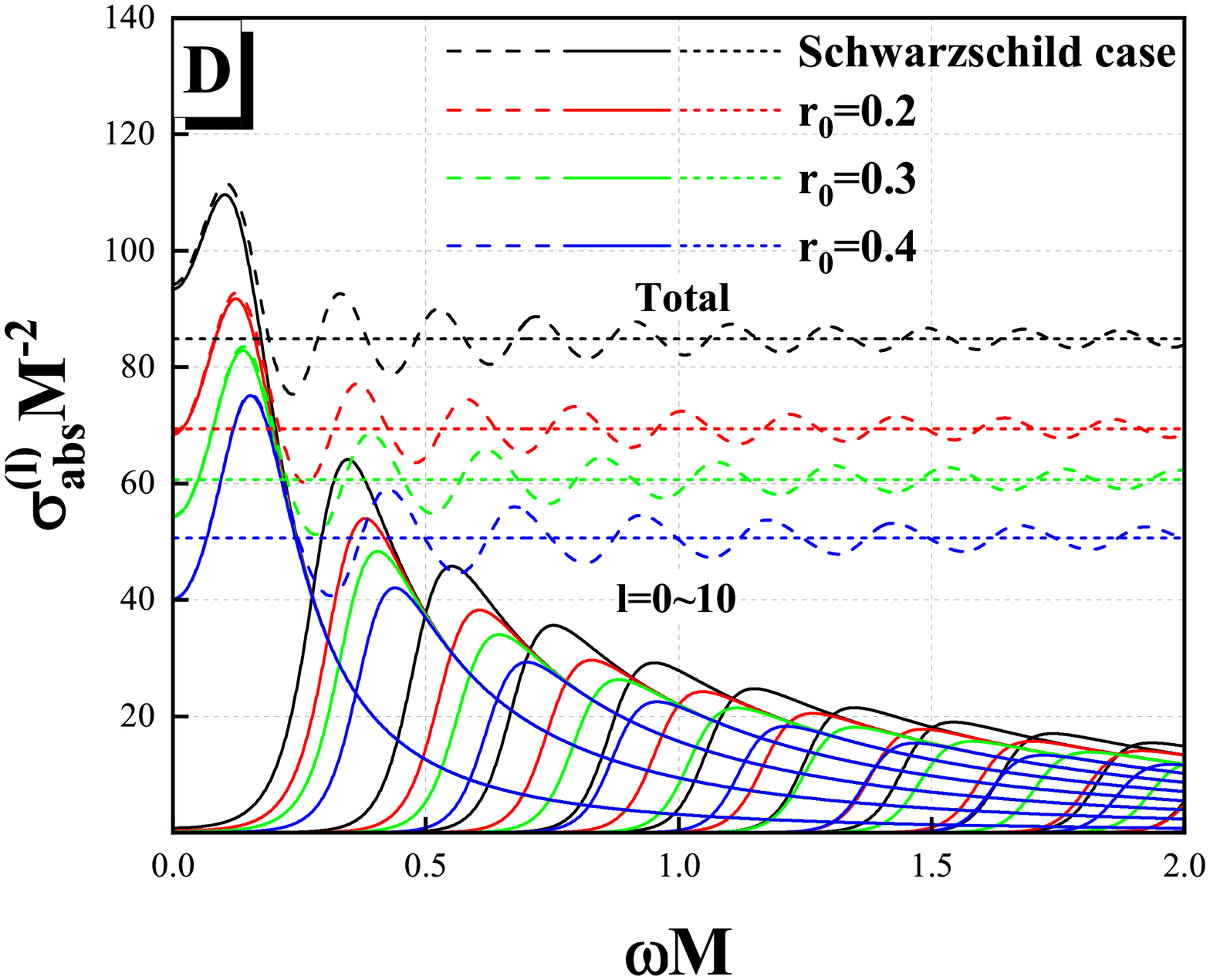}
\includegraphics[width=0.3\textwidth]{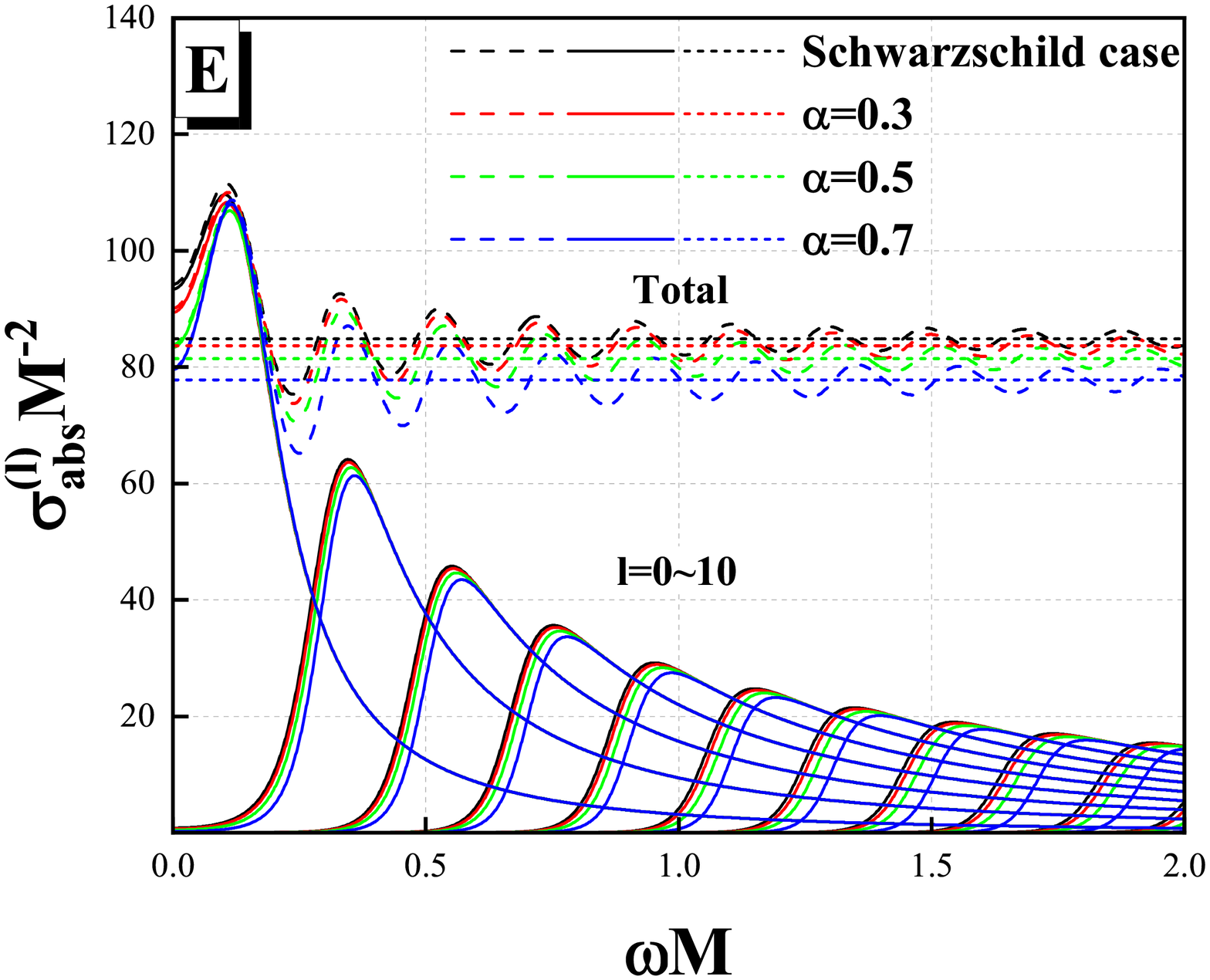}
\includegraphics[width=0.3\textwidth]{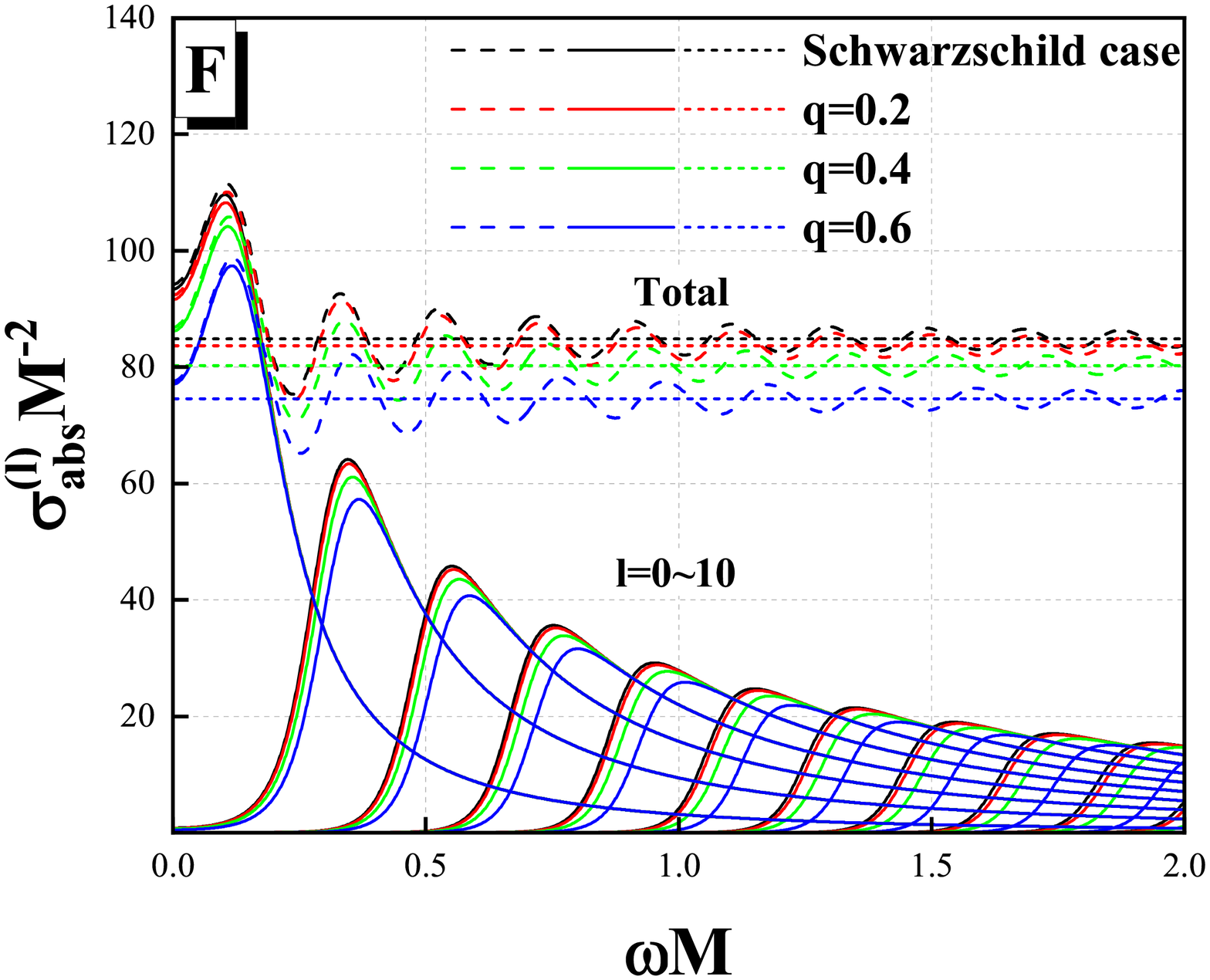}
\caption{
The partial absorption cross sections (solid lines) $\sigma^{(l)}_{\rm{abs}}$, 
i.e. $l=0\sim10$ and total absorption cross sections (dashed lines) $\sigma_{\rm{abs}}$ of the regular black holes with different metric parameters. 
The horizontal lines (short dash) denote the geometrical optics limit value $\sigma^{\rm{hf}}_{\rm{abs}}$. 
The mass $M$ is normalized to 1. 
The letters A-F represent different black holes respectively in Table \ref{tab1}.
} \label{fig3}
\end{figure}

In Fig. \ref{fig3}, 
we show the partial absorption cross sections $\sigma^{(l)}_{\rm{abs}}$, 
i.e. $l=0\sim10$ and total absorption cross sections $\sigma_{\rm{abs}}$ of the regular black holes with different metric parameters. 
The mass $M$ is normalized to 1. 
It is obvious that the partial absorption cross section $\sigma^{(l)}_{\rm{abs}}$ tends to vanish as $\omega M$ increases. 
In the null energy limit, 
the partial wave with $l=0$ provides a nonzero absorption cross section \cite{bib31}. 
Furthermore, 
for each value of $l>0$, the corresponding partial absorption cross section $\sigma^{(l)}_{\rm{abs}}$ starts from zero, 
reaches a peak, 
and decreases asymptotically. 
The larger the value of $l$ is, 
(i)the smaller the corresponding maximum of $\sigma^{(l)}_{\rm{abs}}$ is and (ii) the larger the value of $\omega M$ associated with the peak of $\sigma^{(l)}_{\rm{abs}}$ is. 
This is compatible with the fact that the larger the value of $l$ is, 
the higher the scattering potential $V_{\rm{eff}}$ is.

Fig. \ref{fig3} also shows that the total absorption cross section $\sigma_{\rm{abs}}$ is smaller for bigger value of the metric parameter. 
The total absorption cross section $\sigma_{\rm{abs}}$ oscillates around the limit of geometrical optics. 
When $\omega M\gg1$ goes to the high frequency zone, 
the total absorption cross section tends to the geometric optics limit value $\sigma^{\rm{hf}}_{\rm{abs}}=\pi b^{2}_{\rm{cr}}$ \cite{bib31}, 
where $b_{\rm{cr}}$ is the critical impact parameter mentioned in subsection \ref{subsec2.2}. 
From Fig. \ref{fig1}, 
we can see that the larger the value of the metric parameter is, 
the smaller the critical impact parameter is, 
which is numerically consistent with the results we discussed above.

\begin{figure}[t]
\centering
\includegraphics[width=0.6\textwidth]{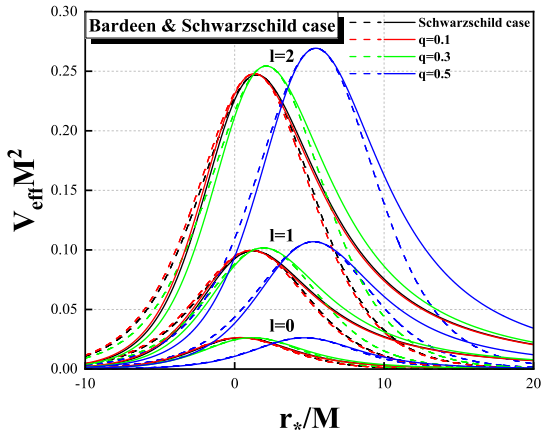}
\caption{
The effective scattering potentials $V_{\rm{eff}}$ as a function of $r_{*}$ for different choices of $l=0 ,1 ,2$ in the background of Bardeen case (red, green and blue lines) and Schwarzschild case (black lines). 
The dashed lines represent the Mashhoon method we utilized and the solid lines stand for the results of numerical integration.
} \label{fig4}
\end{figure}

The absorption of planar massless scalar waves by Bardeen regular black holes has been studied in \cite{bib37}. 
In the low-frequency regime, 
the total absorption cross section is less than the Shwarzschild case in the Figure 6 of \cite{bib37} and is similar as shown in the Fig. \ref{fig3}A of this paper. 
While in the high-frequency regime, 
the trend of the total absorption cross section decreasing with parameter increasing is consistent with Fig. 6 of \cite{bib37} and the oscillation of the total absorption cross section tends to geometrical optics limit $\sigma^{\rm{hf}}_{\rm{abs}}$ in Fig. \ref{fig3}A. 
Because of the Mashhoon method we utilized. 
It can be observed that the accuracy of our estimate increases as the shape of the effective potential becomes more consistent with the P\"oschl-Teller potential. 
We present the effective potential $V_{\rm{eff}}(r_{*})$ with $l=0, 1, 2$ in Fig. \ref{fig4}.
Despite the fact that this approach is fairly simple to compute, it is clear that not all effective potentials of perturbation fields in black holes accord well with the P\"oschl-Teller potential.

\subsection{Scattering cross section} \label{subsec4.2}

\begin{figure}[t]
\centering
\includegraphics[width=0.3\textwidth]{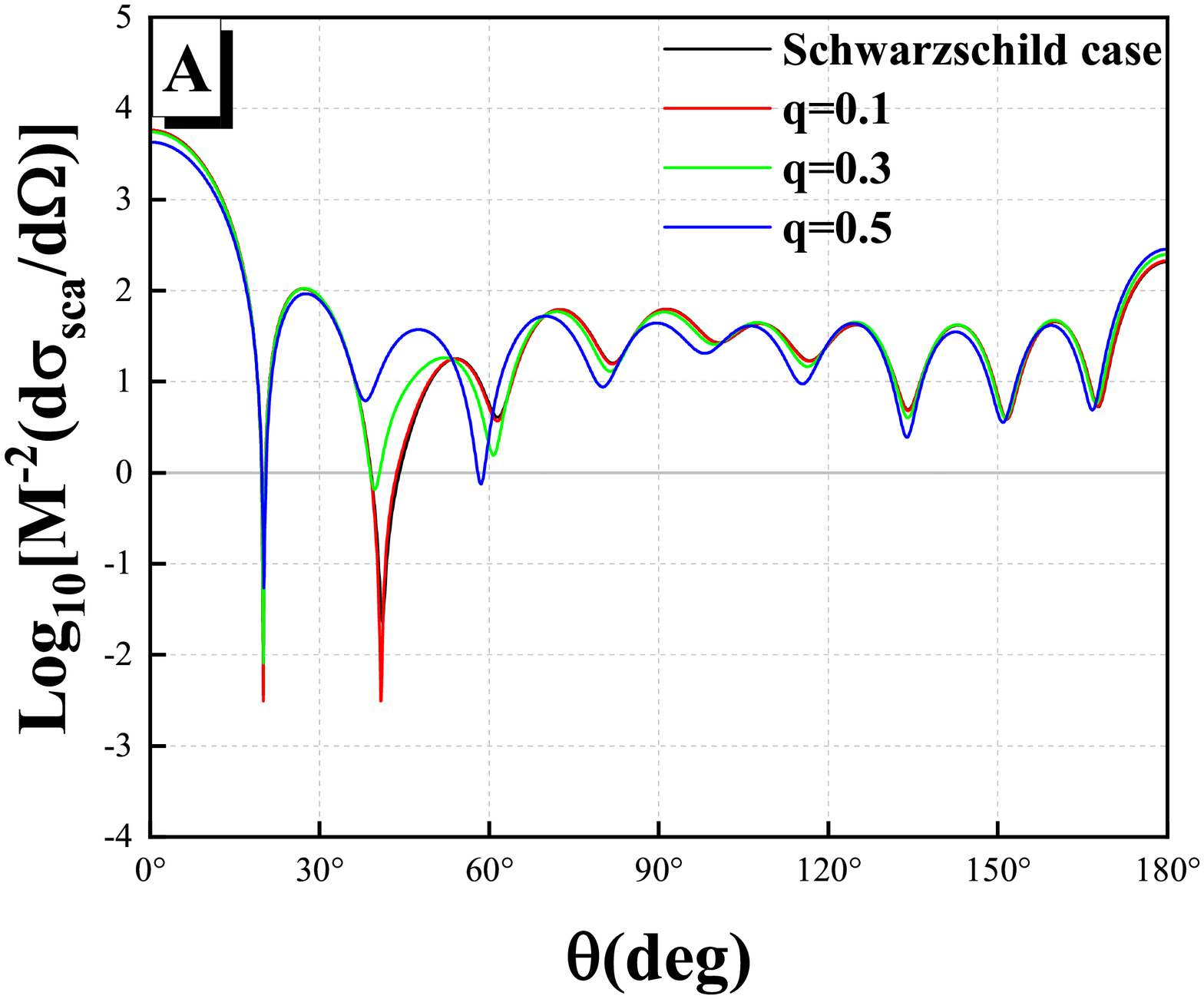}
\includegraphics[width=0.3\textwidth]{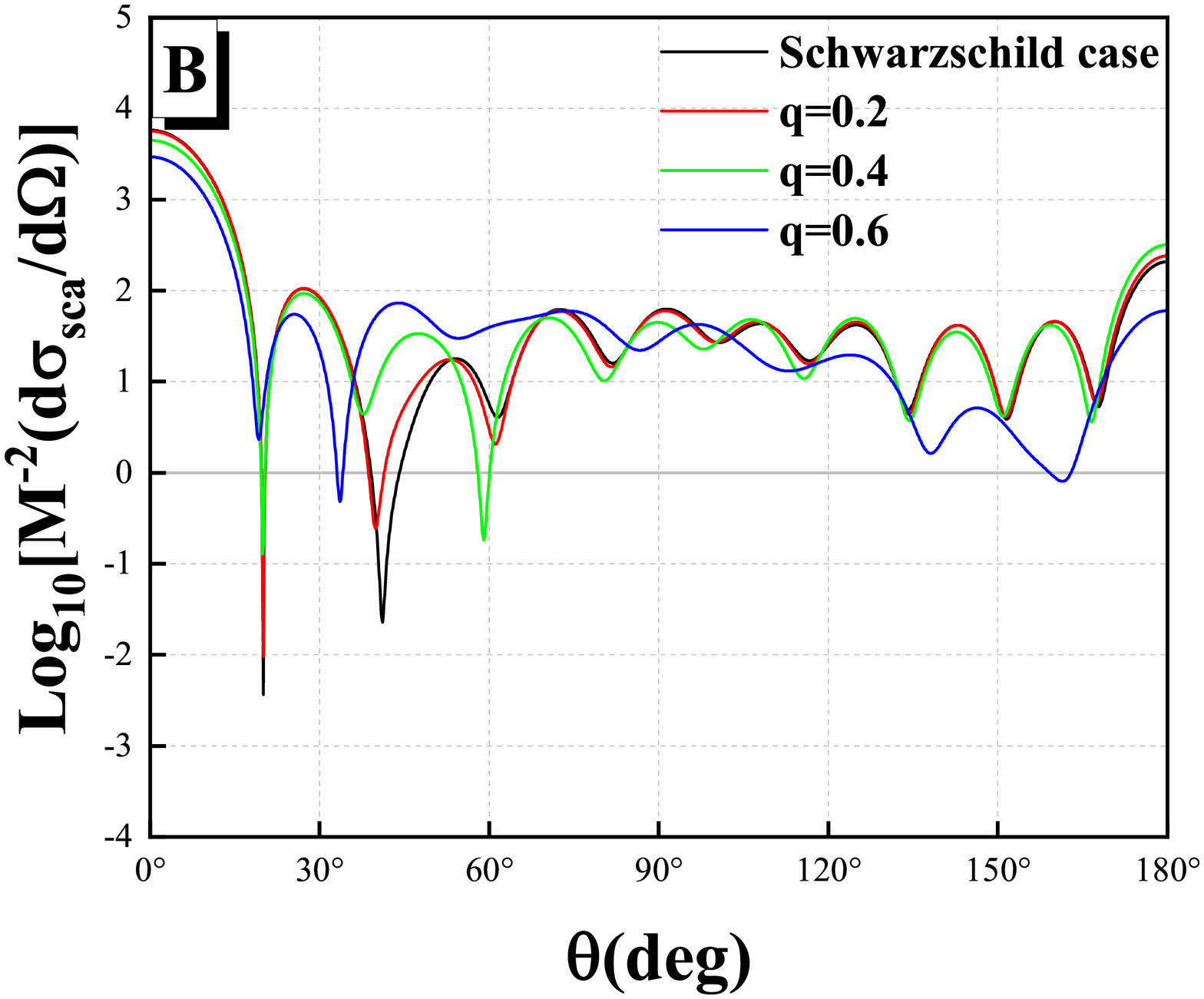}
\includegraphics[width=0.3\textwidth]{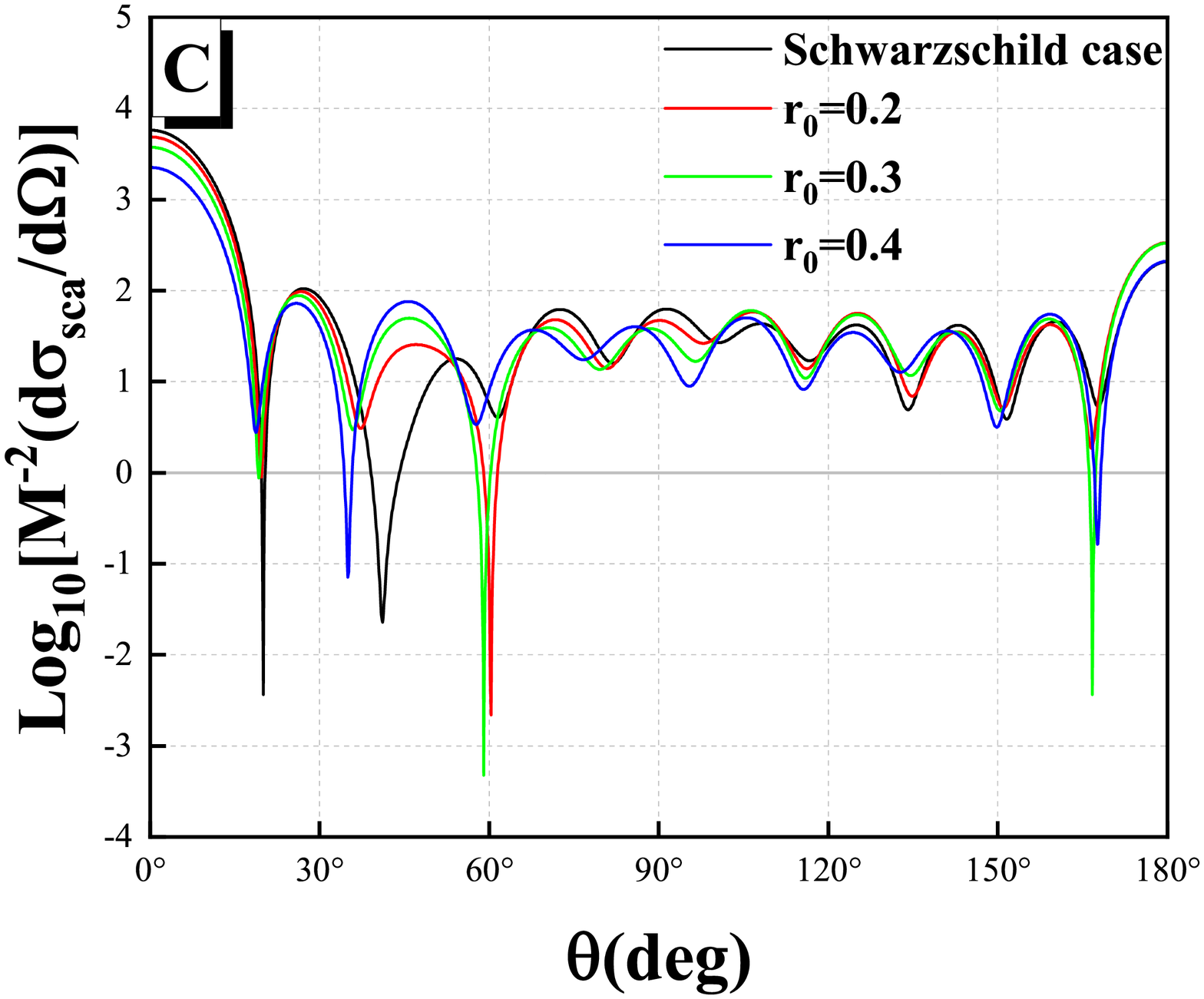}
\includegraphics[width=0.3\textwidth]{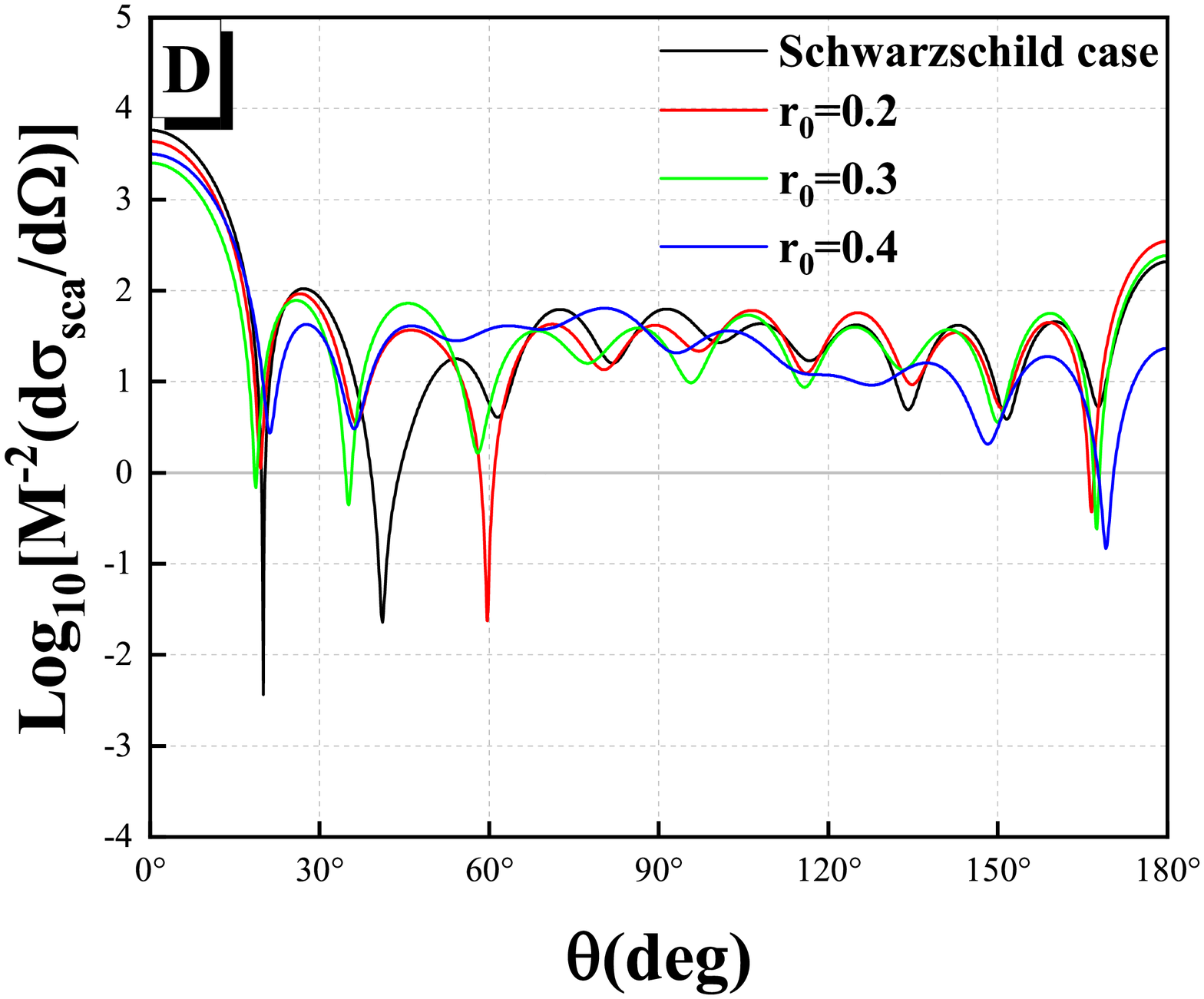}
\includegraphics[width=0.3\textwidth]{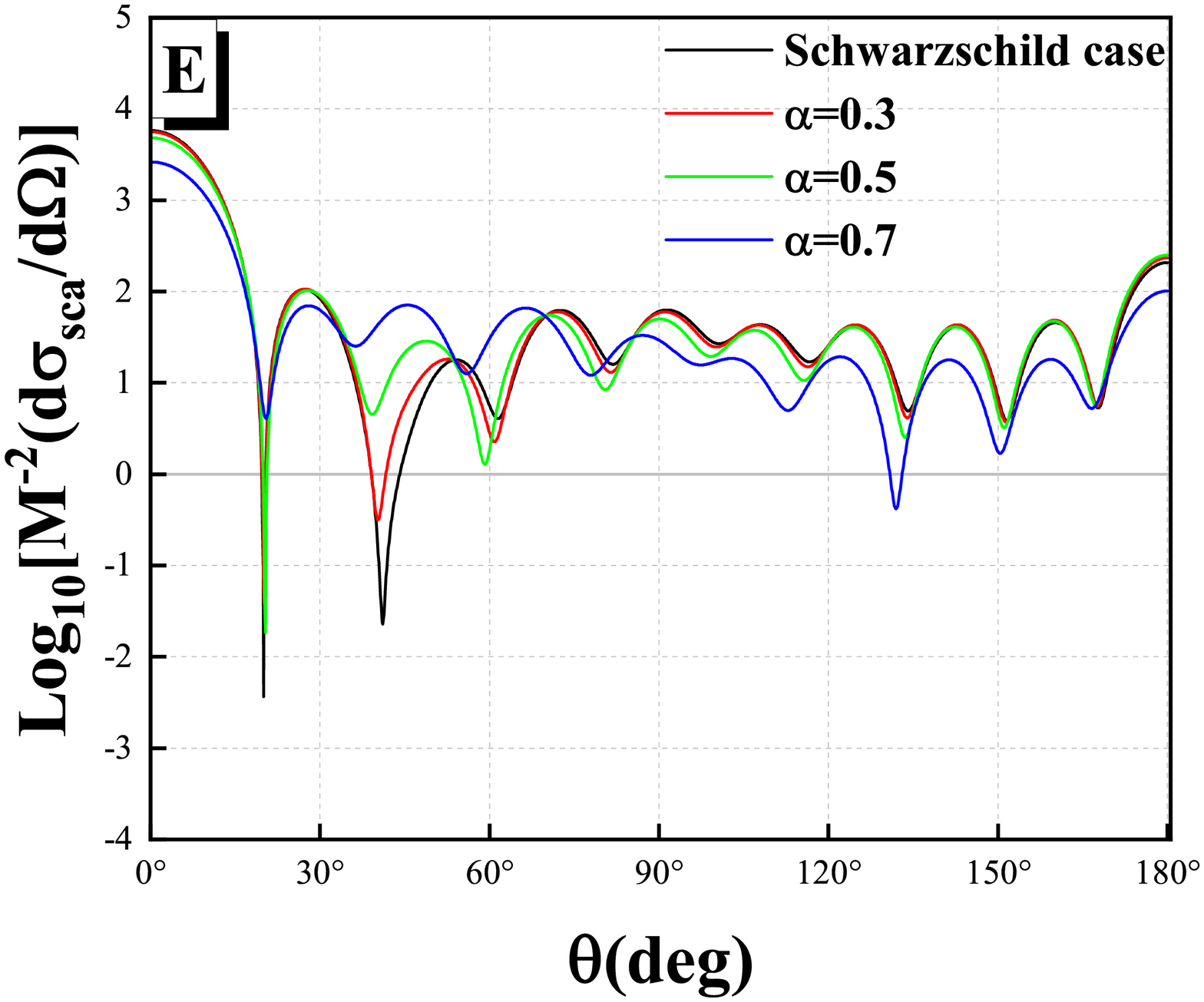}
\includegraphics[width=0.3\textwidth]{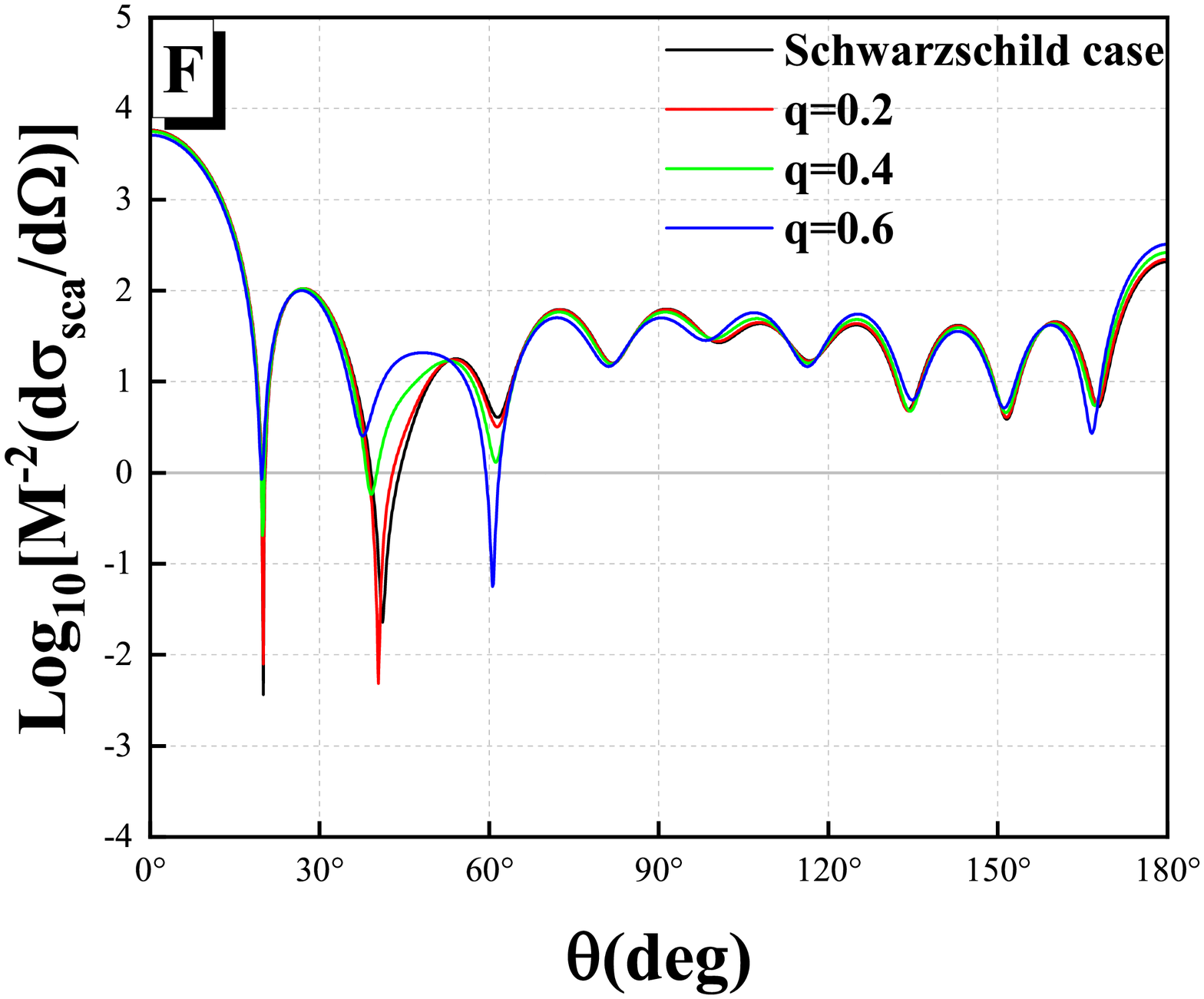}
\caption{
Logarithmic value of the total scattering cross sections of different regular black holes with spacetime parameters in the range of $\theta=0^{\circ}$ to $\theta=180^{\circ}$. 
$\omega M$ is fixed to 1. 
The letters A-F represent different black holes respectively in Table \ref{tab1}.
} \label{fig5}
\end{figure}

It is widely known that the scattering amplitude is defined as
\begin{equation} \label{eq29}
g(\theta)=\frac{1}{2i\omega}\sum\limits_{l=0}^{\infty}(2l+1)(e^{2i\delta_{l}}-1)P_{l}(\cos\theta)
\end{equation}
and the differential scattering cross section
\begin{equation} \label{eq30}
\frac{\mathrm{d}\sigma_{\text{sca}}}{\mathrm{d}\Omega}=\lvert g(\theta)\rvert^{2}
\end{equation}
As a direct consequence, 
the scattering cross section is obtained
\begin{equation} \label{eq31}
\sigma_{\text{sca}}(\omega)=\int\frac{\mathrm{d}\sigma}{\mathrm{d}\Omega}\mathrm{d}\Omega=\frac{1}{2i\omega}\sum\limits_{l=0}^{\infty}(2l+1)\lvert e^{2i\delta_{l}}-1\rvert^{2}
\end{equation}

Fig. \ref{fig5} shows the total scattering cross section as a function of the scattering angle. 
For comparison, 
we also plot the results of the Schwarzschild black hole. 
For some black holes, 
we find the scattering flux is strengthened and its width becomes narrower in the forward direction. 
The scalar field scattering becomes more concentrated. 
With fixed frequency in the low frequency range, 
the glory peak is higher and the glory width becomes narrower. 
As a result, 
the glory phenomena at the forward and backward lends itself advantageously to astronomical observation.

\section{Conclusions} \label{sec5}

In summary, 
we have investigated the scattering and absorption cross section of the massless scalar field from some well-known regular black holes by using the Numerov approach and the P\"oschl-Teller potential approximation. 
Our computational results indicated that the larger the spacetime parameter is, 
the lower the associated total absorption cross section maxima is. 
When compared to the Schwarzschild black hole, 
the scattering cross section is enhanced in some regular black hole spacetimes, 
meanwhile the scattering width is narrow in the forward orientation. 
We also studied the null geodesics in the regular black hole spacetimes and compared the critical impact parameter with the geometrical optical limit. 
The total absorption cross section tends to the geometrical optical limit in the high frequency region. 
We found that in several cases the scattering flux is strengthened and scattering cross section width becomes narrower in the forward direction in comparison to the Schwarzschild case.

\bmhead{Acknowledgments}

The authors thank Dr. Z.N. Yan for his positive help and useful discussion. 
This work was supported by the National Key Research and Develop Program of China under Contract No. 2018YFA0404404.

\bmhead{Data Availability Statement}
The data generated in this study are available on request from the corresponding authors.

\bibliography{RegularBHwithScaler}

\end{document}